\documentclass[lettersize,journal]{IEEEtran}

\usepackage{amsmath,amssymb,amsfonts}
\usepackage{graphicx}
\usepackage{textcomp}
\usepackage{algorithm}
\usepackage{algpseudocode}
\usepackage{gensymb,hyperref}

\usepackage{array}
\usepackage{textcomp}
\usepackage{stfloats}
\usepackage{url}
\usepackage{verbatim}
\usepackage{graphicx}
\usepackage{cite}
\usepackage{rotating}
\usepackage{theorem}
\usepackage{euscript}
\usepackage{empheq}
\usepackage{caption}
\usepackage{subcaption}
\usepackage{xcolor}

\renewcommand{\leq}{\ensuremath{\leqslant}}
\renewcommand{\geq}{\ensuremath{\geqslant}}
\renewcommand{\le}{\ensuremath{\leqslant}}

\newcommand{\argmin}[2]{\ensuremath{\underset{\substack{{#1}}}
{\text{\rm argmin}}\;\;#2}}

\newcommand{\scal}[2]{{\left\langle{{#1}\mid{#2}}\right\rangle}}

\newcommand{\Id}{\ensuremath{\operatorname{Id}}\,}
\newcommand{\RR}{\ensuremath{\mathbb{R}}}

\newcommand{\RPP}{\ensuremath{\left]0,+\infty\right[}}

\newcommand{\NN}{\ensuremath{\mathbb N}}

\newcommand{\proj}{\ensuremath{\text{\rm proj}}}
\newcommand{\prox}{\ensuremath{\text{\rm prox}}}

\def\xb{{\mathbf x}}
\def\yb{{\mathbf y}}

\def\zb{{\mathbf z}}
\def\ub{{\mathbf u}}

\def\sb{{\mathbf s}}
\def\nb{{\mathbf n}}

\def\wb{{\mathbf w}}

\def\Ab{{\mathbf A}}
\def\Bb{{\mathbf B}}
\def\Cb{{\mathbf C}}
\def\Hb{{\mathbf H}}

\def\Rb{{\mathbf R}}
\def\Qb{{\mathbf Q}}

\def\Vb{{\mathbf V}}

\def\Lb{{\mathbf L}}
\def\Wb{{\mathbf W}}
\def\Db{{\mathbf D}}
\def\Fb{{\mathbf F}}
\def\Tb{{\mathbf T}}
\def\Mb{{\mathbf M}}

\begin{document}

\title{Deep Unfolding of the DBFB Algorithm with Application to ROI CT Imaging with Limited Angular Density}
\author{Marion Savanier, Emilie Chouzenoux, Jean-Christophe Pesquet, and Cyril~Riddell
\thanks{This work was supported by the European
Research Council Starting Grant MAJORIS ERC-2019-STG-850925, the ANRT CIFRE Convention 2018/1587, and the ANR Research and Teaching Chair in Artificial Intelligence BRIDGEABLE.}
\thanks{E. Chouzenoux, J.-C. Pesquet and M. Savanier are with Univ. Paris-Saclay, CentraleSup\'elec, CVN, Inria, Gif-sur-Yvette, France (e-mail: firstname.lastname@centralesupelec.fr).}
\thanks{M. Savanier and C. Riddell are with GE Healthcare, Buc, France (e-mail: marion.savanier@gmail.com, cyril.riddell@ge.com).}}

\maketitle

\begin{abstract}
This paper presents a new method for reconstructing regions of interest (ROI) from a limited number of computed tomography (CT) measurements. {Classical model-based iterative reconstruction methods lead to images with predictable features. Still, they often suffer from tedious parameterization and slow convergence. On the contrary,} deep learning methods are fast, and they can reach high reconstruction quality by leveraging information from large datasets, but they lack interpretability. 
At the crossroads of both methods, deep unfolding networks have been recently proposed. Their design includes the physics of the imaging system and the steps of an iterative optimization algorithm. Motivated by the success of these networks for various applications, we introduce an unfolding neural network called U-RDBFB designed for ROI CT reconstruction from limited data. Few-view truncated data are {effectively} handled thanks to a robust non-convex data fidelity term combined with a sparsity-inducing regularization function. We unfold the Dual Block {coordinate} Forward-Backward (DBFB) algorithm, embedded in an iterative reweighted scheme, allowing the learning of key parameters in a supervised manner. Our experiments show an improvement over several state-of-the-art methods, including a model-based iterative scheme, {a multi-scale deep learning architecture}, and other deep unfolding methods.  
\end{abstract}


\section{Introduction}
\label{sec:introduction}

CT imaging is commonly used for diagnostic purposes and image guidance in interventional radiology and surgery. 
In interventions such as follow-up examinations of deployed stents and needle biopsies, only a small region of the patient is of interest.
Irradiating only the ROI by X-rays involves focusing the X-ray beam with collimation techniques using radio-opaque blades before passing through the patient \cite{chityala2004region}. A focused irradiation results in a substantial reduction in patient dose \cite{maier2013new} and truncated measurements (or projections). 
The inverse problem of reconstructing an ROI from a set of truncated projections is ill-posed. In practice, it is often combined with angular sub-sampling for fast or further dose-saving acquisitions. \\ 
Previous research demonstrated the ability of model-based iterative reconstruction (MBIR) for CT reconstruction from few-view measurements \cite{sidky2008image, Bian2010}. MBIR implements nonlinear iterative algorithms aiming at minimizing a penalized cost function. In CT, most works rely on total variation (TV) regularization \cite{Rudin1992}. Although these works achieve a reduction in angular sub-sampling artifacts compared to analytical reconstruction methods, truncated projections still challenge MBIR~\cite{Langet2015}, {especially in terms of computation time}.\\
To implement MBIR, one must choose the reconstruction grid, i.e., the support of the reconstructed area. When the grid matches the support of the ROI, the data will not agree with the reprojection of the ROI. 
When the reconstruction grid includes the support of the entire object \cite{yu2009compressed}, the reconstruction becomes computationally expensive and less stable due to the increase of unknowns for the same amount of data. Truncated data only allows a rough estimation of the exterior anatomical background, which holds no clinical value.
Thus, one more practical solution is to consider an intermediate smaller grid size with a "margin" outside of the ROI \cite{Paleo2017}. This achieves a faster and more stable ROI reconstruction in general. {Yet, when dense objects such as metallic cables or needles are outside the reconstruction grid, the reprojection of the extended ROI contains high-frequency errors. Moreover, when too few projections are used for reconstruction, such objects suffer from aliasing, and additional streak artifacts can degrade the reconstructed ROI~\cite{Langet2015}.} 
Another approach is to reconstruct the entire object with large voxels and then subtract the reprojection of the exterior from the data before reconstructing only the ROI from the subtracted data \cite{Hamelin10, Ziegler08}. 
This approach produces a low-frequency approximation of the exterior of the ROI. {However, such an approximation is poor in the presence of dense objects in the exterior of the ROI, and unwanted high-frequency content remains after subtraction that must again be dealt with.} 
\\
{Apart from the choice for the reconstruction grid}, a way of reducing reconstruction time in MBIR is through preconditioning techniques. However, these techniques are often too restrictive for CT reconstruction when proximal algorithms are used \cite{Savanier2022UnmatchedPO}; hence the emergence of heuristics to accelerate TV-based methods. In \cite{Langet2015}, for example, substantial acceleration of the forward-backward algorithm is reached thanks to two modifications. The first modification involves weighing a least-squares data fidelity term by the ramp filter. Each gradient step then performs an approximate inversion of the forward projection, and each proximity step acts as a post-processing filter. The use of an approximate inverse of the measurement operator in the data fidelity term is also advocated in \cite{Tirer2021, Tirer2020, nilchian2013iterative}. 
The second modification is the linear decrease of the regularization parameter strength along with the iterations. 
Despite these modifications, the proximity operator of TV regularization does not have a closed form, so the method still requires many sub-iterations. Thus, designing MBIR methods with a low computational cost for CT imaging is still challenging.
\\ 
An increasingly popular alternative to MBIR is convolutional neural networks (CNN) due to their increased expressivity and fast inference. CNNs, and in particular U-net \cite{Ronneberger2015}, have already been used for removing sub-sampling streaks in CT reconstructions obtained using analytical methods \cite{Han2016, Jin2017}. 
However, there are concerns about the lack of guarantees and capacity for generalization of post-processing CNNs, because these networks do not ensure data consistency \cite{Sidky2020}. 
The deep unfolding paradigm \cite{monga2021algorithm, wang2020deep} circumvents this issue by offering a way to include a priori information in a neural network. Unfolding networks have been applied to many inverse problems, such as denoising \cite{Wang2016, Pustelnik2022}, deblurring \cite{Bertocchi2020}, MRI reconstruction \cite{Yang2016}, and CT reconstruction from few-view data \cite{Adler2017}. Unfolding consists of untying each iteration of an optimization algorithm for MBIR, defining a set of learnable parameters, and training each iteration (or layer) in an end-to-end manner. 
Some authors allow the learning of the optimization algorithm hyperparameters \cite{Bertocchi2020} as well as linear operators in the regularization, such as convolution kernels in ISTA-net \cite{Zhang2018} and in \cite{Pustelnik16}. Others use CNNs to replace proximity operators, as in PD-net \cite{Adler2017} and ADMM-net \cite{Yang2016}. Deep unfolding networks automatically inherit from the feedback mechanism of MBIR for data consistency.
\\
In this paper, we design a deep unfolding network for addressing the problem of ROI image reconstruction from few-view truncated measurements.
Specifically:
\begin{itemize}
    \item We first introduce an original cost function involving a Cauchy fidelity term and a semi-local total-variation regularization to limit sub-sampling streaks from objects inside and outside the reconstruction grid.
    \item We then advocate for an iterative optimization algorithm combining for the first time an instance of the dual block forward-backward algorithm (DBFB) \cite{Abboud17} with an iterative reweighted scheme.
    \item We propose a neural network architecture inspired by this algorithm, allowing supervised parameter learning and fast reconstruction on the GPU.
    \item We evaluate the performance and generalization of our network on three datasets against state-of-the-art methods such as MBIR, other deep unfolding networks, and classical deep learning post-processing. 
\end{itemize}
This paper is organized as follows: Section~\ref{sec:Notation} introduces our notation, and Section~\ref{sec:Pb} provides a mathematical formulation of our reconstruction problem. Section \ref{sec:theory} presents our cost function and introduces a convergent iterative algorithm to minimize it. Section \ref{sec:unfold} then explains how this algorithm is unfolded into a deep learning architecture and discusses our training strategy. This is followed by experiments (Section~\ref{sec:exp}), results (Section \ref{sec:res}), and discussions (Section \ref{sec:disc}).

\section{Notation}\label{sec:Notation}
 
Throughout the paper, the underlying image space is the Euclidean space $\RR^L$ equipped with the standard scalar product $\langle \cdot,\cdot \rangle$, and the norm $\|\cdot\|$. Moreover, $|||\Lb|||$ denotes the spectral norm of squared matrix 
$\Lb \in \RR^{L \times L}$. Let $\mathcal{S}^{+}_L$ be the set of symmetric positive definite matrices in $\RR^{L \times L}$. For $\Qb \in \mathcal{S}^+_L$, $\| \cdot \|_{\Qb}$ denotes the $\Qb$-weighted norm, i.e., for every $\xb \in \RR^L$,  $\|\xb\|_{\Qb} = \sqrt{\scal{\xb}{\Qb \xb}}$. 
The diagonal matrix with diagonal entries equal to vector $\zb = (z_l)_{\ell=1}^L \in \RR^L$ is denoted $\operatorname{diag}((z_\ell)_{\ell=1}^L)$.
The class of functions which are proper, convex, lower-semicontinuous, defined on $\RR^L$ and taking values in $\RR \cup \{+\infty\}$ is denoted by $\Gamma_{0}(\RR^L)$. The proximity operator of $g \in \Gamma_{0}(\RR^L)$ at $\xb \in \RR^L$
is uniquely defined as \cite{Livre1}
   $\prox_{g}(\xb) = \argmin{\zb \in \RR^L}{\left( g(\zb ) + \frac{1}{2} \|\xb-\zb \|^2  \right)}$.
The indicator function of a nonempty subset $C$ of $\RR^L$ is the function $\iota_C$ equal to 0 on $C$
and $+\infty$ outside of $C$. If $C$ is closed and convex, the projection onto $C$
is denoted by $\proj_C$.

\section{Problem formulation}\label{sec:Pb}

We consider a 1D detector array of $B \in \NN$ bins rotating around an object. The detector is too short to measure the projections of the entire object; its size defines a circular ROI we aim to reconstruct.
Let $S\in \NN$ be the number of projection angles. The vector of truncated sub-sampled tomographic data is $\yb \in \mathbb{R}^{T}$ with $T = B\, S$.\\
A reconstruction of the object attenuation map in the ROI can be obtained by considering a model of the form: 
\begin{equation}\label{eq:decomp_p}
 \Hb \overline{\xb}_{\rm G} = \yb + \nb,
\end{equation}
where $\nb \in \mathbb{R}^{T}$ accounts for some acquisition noise, $\overline{\xb}_{\rm G} \in \RR^{L}$ {is the scanned image restricted to a grid $G$ with support larger than the ROI but smaller than that of the entire object}, and $\Hb \in  \mathbb{R}^{T \times L}$ is the projector that models projection over this intermediary grid $G$.
Operator $\Hb$ contains a subset of the columns of the projector on the entire space, or equivalently it corresponds to setting a subset of the columns of the full projector, corresponding to the pixels outside of $G$, to zero.\\
When the grid $G$ corresponds to the ROI, \eqref{eq:decomp_p} assumes that the image values are 0 outside of the ROI. This assumption is not necessarily valid for truncated data, so \eqref{eq:decomp_p} does not hold. 
Hereafter, we suppose that the grid is extended beyond the ROI so that no assumption is made about the values outside the ROI. 
\\
{MBIR} finds an estimate of $\overline{\xb}_{\rm G}$ by computing a minimizer of a penalized cost function that can, for instance, be written as 
the sum of a data fidelity term $f$ involving $\Hb$ and $\yb$, and a regularization term $r$, as
\begin{equation}\label{eq: generic_cost_func}
    \argmin{\xb \in \RR^L}{f(\xb) + r(\xb)}.
\end{equation}
{Our proposed neural network will unfold a baseline algorithm for solving the MBIR problem \eqref{eq: generic_cost_func} for a specific choice of $f$ and $r$. The next section presents the two terms of the cost function and the baseline nested-loop minimization algorithm.
}

\section{{Model-based} iterative reconstruction}\label{sec:theory}

\subsection{Cost function}\label{subsec:cost_func}

\subsubsection{Cauchy data fidelity}

\begin{figure}[h!]
\centering
    \includegraphics[width=0.45\textwidth]{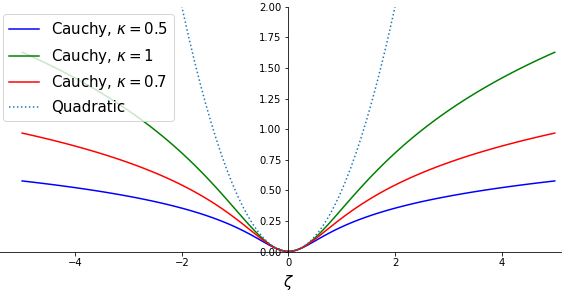}
    \caption{Comparison between the Cauchy and the quadratic functions.} 
    \label{fig:Cauchy}
\end{figure}

When objects with high gradients (metallic wires or needles) do not belong to the reconstruction grid $G$, the error between the data and the reprojection of the estimate over $G$ contains outliers at the projections of those high gradients. Angular sub-sampling of these outliers leads to streaks originating from these objects, i.e., from outside the grid. This means the data should not be trusted equally but through a statistical analysis different from measurement noise.
To avoid the streaks, we propose to decrease the influence of the largest errors between $\yb$ and the reprojection of $\Hb \overline{\xb}_{\rm G}$ using an M-estimator. \\
Here, we focus on the Cauchy estimator $\phi$, which is a redescending M-estimator \cite{andrews2015robust} i.e., its derivative decreases to zero on $]-\infty,\kappa] \cap [\kappa,+\infty[$. It is defined as
\begin{equation}\label{eq:Cauchy_func}
    (\forall \zeta \in \RR) \qquad \phi(\zeta) = \frac{\beta \kappa^2}{2} \ln \left(1+\left(\frac{\zeta}{\kappa}\right)^2 \right),
\end{equation}
where $\beta >0$ is a weighting term and $\kappa>0$ monitors the sensitivity to outliers: the lower $\kappa$, the lower the influence of the outliers. \\
A graphical comparison of the Cauchy function \eqref{eq:Cauchy_func} and the quadratic function $\phi(\cdot) = \frac{\beta}{2}(\cdot)^2$  is displayed in Figure~\ref{fig:Cauchy} for $\beta=1$ and various values of $\kappa$.
\\
Our data fidelity term $f$ reads
\begin{equation}
(\forall \xb \in \RR^L) \quad
    f(\xb) = g(\Hb \xb - \yb),
\end{equation}
with 
\begin{equation}
  (\forall \zb = (z_t)_{1\le t\le T} \in \RR^{T}) \qquad  g(\zb)= \sum^{T}_{t=1}\phi(z_t).
\end{equation}

\subsubsection{Semi-local total variation}

\begin{figure}[h!]
\centering
\includegraphics[width=3cm]{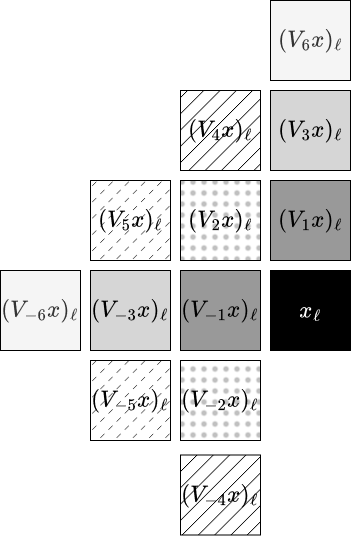}
\caption{Shift operators $(V_j)_{j \in \Lambda_6}$ applied to a given pixel position~$\ell$}
\label{fig:neighborhood}
\end{figure}

Sub-sampling streaks are commonly handled by total variation (TV) regularization \cite{Rudin1992, sidky2008image, Langet2015}. 
Semi-local variants (STV)~\cite{Kindermann06} extend TV in a neighborhood of pixels indexed in $\Lambda_J = \{-J, \ldots, J\}\setminus \{0\}$:
\begin{align}\label{eq:STV}
(\forall \xb \in \RR^L) & \nonumber\\
    r_{\rm STV}(\xb) &= 
      \sum_{j=1}^J \sum_{\ell=1}^L \alpha_{j,\ell} \sqrt{(\xb - \Vb_j \xb)_\ell^2 + (\xb- \Vb_{-j} \xb)_\ell^2} \nonumber\\
    &= \sum_{j = 1}^J r_j(\nabla_j \xb).
\end{align}
Hereinabove $\ell\in \{1,\ldots,L\}$ is the  spatial index and
$\Vb_j, \Vb_{-j} \in \RR^{L \times L}$ are shift operators as shown in Figure~\ref{fig:neighborhood} for $j \in \{1, \ldots, J\}$ and $J=6$. Moreover, for every $j \in \{1, \cdots, J\}$, we define $\nabla_j = \begin{bmatrix} \Vb_j^\top & \Vb_{-j}^\top \end{bmatrix}^\top \in \RR^{2L \times L}$ and, for every $\zb = (\zb_1,\zb_2) \in \RR^{2L}$, $r_j(\zb) = \sum_{\ell=1}^L \alpha_{j,\ell} \sqrt{(\zb_{1})_\ell^2 + (\zb_{2})_\ell^2}$. \\Parameters 
$(\alpha_{j,\ell})_{1\le j \le J,1\le \ell \le L}$ are nonnegative weights that can be chosen to vary spatially, so making STV adaptive to the spatial contents \cite{gilboa2009nonlocal}. We recover the standard TV regularization for constant values of these parameters and $J=1$.
\\
We add a nonnegativity constraint on the pixel values and a quadratic term $\frac{1}{2} \|\xb\|^2_{\Mb} = \xb^\top \Mb \xb$ to the STV regularization. Here, matrix $\Mb = \operatorname{diag}((m_{\ell})^{L}_{\ell=1}) \in \mathcal{S}_L^+$ is such that, for 
every $\ell \in \{1,\ldots,L\}$, 
$m_{\ell} = 1$ if the $\ell$-th entry 
$x_{\ell}$ of vector $\xb$ belongs to the ROI, and $m_{\ell} = \xi > 1$ otherwise. Thus $\Mb \in \mathcal{S}_L^{+}$ acts as a mask, limiting high values outside of the ROI. \\
\\
Altogether, our regularization function in \eqref{eq: generic_cost_func} reads
\begin{equation}\label{eq:expression_r}
(\forall \xb \in \RR^L) \quad
    r(\xb) =  \sum_{j=1}^J r_j(\nabla_j \xb) + \frac{1}{2} \|\xb\|^2_{\Mb} + \iota_{[0,+\infty[^{L}}(\xb).
\end{equation}

To our knowledge, STV regularization has not been used in CT for handling sub-sampling streaks. In this work, STV provides extra capacity compared to TV for learning. The Cauchy fidelity term has been used in ultrasound imaging \cite{Ouzir2019} and in CT imaging \cite{Kazantsev2017} for mitigating the ring artifacts that appear due to defective detector bins only.

\subsection{Minimization algorithm}

\begin{figure*}[t]
\centering
\includegraphics[width=18cm]{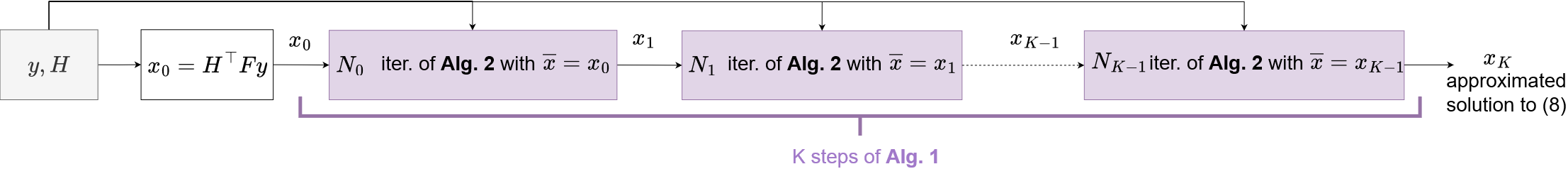}
\caption{{Baseline RDBFB algorithm for solving \eqref{eq:pb0}: $K$ steps of Alg. 1, where each step $k$ consists of applying $N_k$ iterations of Alg. 2.}}
\label{fig:flow}
\end{figure*}

\subsubsection{Reweighting for non-convex data fidelity}
Given our choices for $r$ and $f$, Problem \eqref{eq: generic_cost_func} becomes
\begin{equation}\label{eq:pb0}
    \argmin{\xb \in \RR^L}{g(\Hb\xb - \yb)  + \sum_{j=1}^J r_j(\nabla_j \xb) + \frac{1}{2} \|\xb\|^2_{\Mb} + \iota_{[0,+\infty[^{L}}(\xb)}.
\end{equation}
Because of the non-convexity of $g$, we adopt an iterative reweighting strategy where Problem~\eqref{eq:pb0} is replaced by a sequence of surrogate convex problems built following a majoration principle. As shown in Appendix~\ref{a:MM}, the associated Majorization-Minimization strategy is given by Algorithm~\ref{alg:rewighted}, 
\begin{algorithm}
\caption{Iterative reweighting strategy for Problem \eqref{eq:pb0}}\label{alg:rewighted}
\begin{algorithmic}
\Require {Number of iterations $K \in \NN^*$, $\xb_0 \in \RR^L$}
\For{$k=0$ to $K-1$}
\State Define majorant function $Q(\xb; \overline{\xb})$ {using \eqref{eq_funcQ}}
\State \begin{equation}\label{eq:step_algo1}
    \xb_{k+1} = \argmin{\xb \in \RR^L}{Q(\xb, \xb_k)}
\end{equation}
\EndFor
\Ensure {$\xb_K$ approximating the solution to \eqref{eq:pb0}}
\end{algorithmic}
\end{algorithm}
\\
where, for every $\overline{\xb} \in \RR^L$, the surrogate function $Q(\cdot,\overline{\xb})$ is 
\begin{equation}
    Q(\xb, \overline{\xb})= \iota_{[0,+\infty[^{L}}(\xb)+h_0(\Bb_0 \xb;\Bb_0 \overline{\xb})  + h_1(\Bb_1 \xb)+ \frac{1}{2}\|\xb\|^2_{\Mb},
    \label{eq_funcQ}
\end{equation}
where $\forall (\xb,\overline{\xb}) \in (\RR^L)^2$,
with $\Bb_0 =  \Hb \in \RR^{T\times L}$,
\begin{align} \label{eq:def_B}
    \Bb_1 &= \begin{bmatrix} \nabla_1^\top  & \cdots & \nabla_j^\top \end{bmatrix}^\top \in \RR^{2JL\times L}\nonumber\\
h_0(\cdot; \Bb_0 \overline{\xb}) &= \tilde{g}(\cdot - \yb; \Bb_0 \overline{\xb} -  \yb)\nonumber\\
h_1(\Bb_1\xb)&=\sum_{j=1}^J r_j(\nabla_j \xb). 
\end{align}
Let $(\xb_k)_{k\in \NN}$ be generated by Algorithm \ref{alg:rewighted}. The 
cost sequence value (defined from \eqref{eq:pb0}) monotonically converges.

\subsubsection{Dual block coordinate forward-backward algorithm}

\begin{algorithm}[h]
\caption{DBFB algorithm to solve \eqref{eq:step_algo1} with ${\xb}_k = \overline{\xb} \in \RR^L$}\label{alg:DBFB}
\begin{algorithmic}
\Require Number of iterations $N \in \NN^*$, tangent point $\overline{\xb} \in\RR^L$, initial dual variables $\zb_0^0 \in \RR^T, (\forall j \in \{1,\ldots,J\}) \; \zb^j_0 \in \RR^{2L}$ with constant $J$ defined in \eqref{eq:STV}, operators $\Bb_0$ and $\Bb_1$ defined in \eqref{eq:def_B}, stepsizes $(\sigma,\tau_1,\ldots,\tau_J)\in ]0,+\infty[^{J+1}$.
\end{algorithmic}
\begin{align}
& \sb^1_0 = ({\zb^1_0}, \ldots,  {\zb^J_0}) \notag\\
& \mathbf{\Sigma} = \operatorname{diag}(\tau_1,\ldots,\tau_J)\notag\\
&\wb_0 = -\Mb^{-1} (\Bb_0^\top \zb_0^0 +\Bb_1^\top \sb_0^1) \notag \\ 
& \text{For }n=0,1\ldots, N \notag\\
&\quad
\left\lfloor
\begin{array}{l}
  ({\zb^1_n}, \ldots, {\zb^J_n}) \equiv \sb^1_n \nonumber \\
\boldsymbol{x}_n = \proj_{[0,+\infty[^{L}}(\wb_n \nonumber)\\
\mbox{Select $\varepsilon_{n} \in \{0,1\}$ and $\gamma_n \in ]0,+\infty[$} \nonumber \\
\text{If }\varepsilon_{n}=0 \quad \text{\textbf{(D)}}\\
\quad\left\lfloor
\begin{array}{l}
     \Tilde{\zb}_n^{0} = \zb_n^{0} + \gamma_n \sigma^{-1} \Bb_0 \boldsymbol{x}_n\nonumber\\
     \zb_{n+1}^{0} = \Tilde{\zb}_n^{0}  - \gamma_n \sigma^{-1} \prox_{\gamma_n^{-1} \sigma h_0(\cdot; \Bb_0 \overline{\xb})}(\gamma_n^{-1} \sigma \Tilde{\zb}_n^{0}) \nonumber \\
     \wb_{n+1} = \wb_n - \Mb^{-1} \Bb_0^\top (\zb_{n+1}^{0} - \zb_n^{0})\\
     \sb_{n+1}^1 = \sb_n^1 
    \end{array}\right. \\
 \text{If }\varepsilon_{n}=1 \quad \text{\textbf{(R)}}\\ 
\quad \left\lfloor
\begin{array}{l}
   \Tilde{\sb}_n^{1} = \sb_n^{1} + \gamma_n \mathbf{\Sigma}^{-1} \Bb_{1} \boldsymbol{x}_n\nonumber\\
  \sb_{n+1}^{1} = \Tilde{\sb}_n^{1} - \gamma_n \mathbf{\Sigma}^{-1} \prox_{\gamma_n^{-1} \mathbf{\Sigma} h_{1}}(\gamma_n^{-1} \mathbf{\Sigma} \Tilde{\sb}_n^{1}) \nonumber\\
  \wb_{n+1} = \wb_n - \Mb^{-1} \Bb_1^\top (\sb^1_{n+1} -\sb^1_n)\\
 \zb_{n+1}^0 = \zb_n^0
    \end{array}\right. \\
\end{array}\right. 
\end{align}
\begin{algorithmic}
\Ensure {$\boldsymbol{x}_N$ approximating the minimizer of $Q(\cdot,\overline{\xb})$.}
\end{algorithmic}
\end{algorithm}

The $k$-th iteration of Algorithm \ref{alg:rewighted} requires to solve \eqref{eq:step_algo1}, which amounts to minimizing the function $Q(\cdot, \overline{\xb})$, with $\overline{\xb}$ equals to the current iterate $\xb_k$.
Since $\Mb \in \mathcal{S}^{+}_L$, $Q(\cdot, \overline{\xb})$ is strongly convex for every $\overline{\xb} \in \RR^L$. The minimization \eqref{eq:step_algo1} is hence well-defined, with a unique solution that can be conveniently obtained using the dual forward-backward algorithm \cite{combettes2011proximal, Pustelnik2022}. 
An accelerated and light version of this algorithm is its block coordinate version (DBFB) \cite{abboud2017dual}, which allows for accessing the proximity operators of $h_0$ and $h_1$ separately. \\Algorithm \ref{alg:DBFB} describes $N \in \NN^*$ iterations of DBFB. The output $\boldsymbol{x}_N$ generated by DBFB with input $\xb_k$ then defines $\xb_{K+1}$ in Algorithm~\ref{alg:rewighted}.
{For every $n \in \{1,\ldots,N\}$, DBFB updates the main primal variable $\boldsymbol{x}_n$ as well as two dual variables $\zb_n^0 \in \RR^{T}$ and $\sb_n^1 \in \RR^{2JL}$, associated with the data fidelity (\textbf{data step (D)}) or the regularization (\textbf{regularization step (R)}) terms, respectively. Each dual variable is activated (or not) at iteration $n$ according to a binary variable $\varepsilon_n$. 
}
\\
When $N \to \infty$, the DBFB sequence $(\boldsymbol{x}_n)_{n \in \mathbb{N}}$ converge to the solution to \eqref{eq:step_algo1}
under the following assumptions on the algorithm parameters~\cite{abboud2017dual}:
\begin{equation*}
    \begin{cases}
    \sigma \geq ||| \Bb_0 \Mb^{-1} \Bb_0^\top |||,\\
    (\forall j \in \{1,\ldots,J\}) \quad \tau_j \geq ||| \nabla_j \Mb^{-1} \nabla_j^\top |||,\\
    \gamma_n \in [\epsilon, 2-\epsilon] \text{ with }    \epsilon \in ]0,1]\\
    (\exists M \in \NN\setminus\{0,1\})(\forall n\in \NN)\quad 
    0<\sum_{n'=n}^{n+M-1} \varepsilon_{n'}<M.
    \end{cases}
\end{equation*}
The first three assumptions are stepsize range conditions.
The last one means that each step \textbf{(D)} and \textbf{(R)} is performed at least once every $M$ iterations.
The practical implementation of the proximal operators involved in DBFB for our choices for $h_0$ and $h_1$ is discussed in Appendix \ref{a:DFB}. 
\\
\\
The overall iterative strategy for approximating the solution to \eqref{eq:pb0} consists of applying Algorithm~\ref{alg:rewighted}, where, for every $k \in \NN^*$, $N_k \in \NN^*$ iterations of Algorithm~\ref{alg:DBFB} are used as an inner solver with $\overline{\xb} = \xb_k$ to compute $\xb_{k+1}$ in \eqref{eq:step_algo1}. We call the resulting iterations \textit{reweighted DBFB (RDBFB) algorithm} {(see Figure \ref{fig:flow})}. 
Such a combination of an iterative reweighted algorithm with dual ascent steps is original to our knowledge. In the context of CT, iterative reweighted algorithms usually involve surrogates to the regularization term~\cite{Yu2017, Wang2017} rather than to the data fidelity, as done here.

\newpage

\section{Unfolded reconstruction}\label{sec:unfold}

{Hereinafter, we present a deep neural network, designated as {U-RDBFB} (Unfolded Reweighted DBFB), by unfolding all the steps of RDBFB. Specifically, the network mimics the application of $K$ iterations of Algorithm \ref{alg:rewighted}, as $K$ main layers, each of them grouping $N_k \in \NN$ iterations of Algorithm \ref{alg:DBFB}. This yields an architecture with $\sum_{k=0}^{K-1} N_k$ layers in total.}

\subsection{From RDBFB iterations to U-RDBFB layers}
{The deep unfolding paradigm recasts every step of Algorithm \ref{alg:DBFB} as one neural network layer: step \textbf{(D)} becomes $\mathcal{L}_D$ ($\varepsilon_n = 0$) and step \textbf{(R)} becomes $\mathcal{L}_R$ ($\varepsilon_n = 1$). It requires truncating the number of layers drastically. To optimize the depth of our network, we propose two modifications of the steps \textbf{(D)} and \textbf{(R)} to construct the corresponding layers.} 
These modifications stem from recent works \cite{Chouzenoux2021, Elfving2018, Lore18, Andres} that have shown the potential to replace adjoint operators with surrogates to accelerate the convergence of MBIR methods \cite{Zeng2019, Zeng2000, Sidky2022}. In CT, a frequently encountered operator is the ramp filter $\Fb$, which satisfies $\Fb \Hb \Hb^\top \approx \Id$. Thus, to improve conditioning and allow larger values for the step sizes associated with layer $\mathcal{L}_D$, hence a lower number of such a layer, we replace $\Bb_0$ in Algorithm~\ref{alg:DBFB} with $\Fb \Hb$. 
\\
By setting $\nu_{n,0}= \gamma_n \sigma^{-1}$ and by using the relation
\[
\prox_{\nu_{n,0} h_0(\cdot; \Fb \Hb \overline{\xb})}
= \prox_{\nu_{n,0} \tilde{g}(\cdot; \Hb \overline{\xb} - \yb)}
(\cdot
-\Fb\yb)+\Fb\yb,
\]
we define layer $\mathcal{L}_{\rm D}$ as 
\begin{align}\label{eq:layerD}
&\mbox{\textbf{Data layer ($\mathcal{L}_{\rm D}$):}}\nonumber\\
&\quad
\left\lfloor
\begin{array}{l}
    \boldsymbol{x}_n = \proj_{[0,+\infty[^{L}}(\wb_n)\\
   \ub_n = \zb_n^{0} + \nu_{n,0} \Fb(\Hb \boldsymbol{x}_n - \yb)\\
    \zb_{n+1}^{0} =  \ub_n- \nu_{n,0}^{-1}\, \prox_{\nu_{n,0} \tilde{g}(\cdot; \Fb \Hb \overline{\xb} - \yb)} (\nu_{n,0}  \ub_n) \\
    \wb_{n+1} = \wb_n - \Mb^{-1} \Hb^\top  (\zb_{n+1}^{0} - \zb_n^{0})\\
    \zb_{n+1}^j = \zb_n^j \quad (\forall j \in \{1,\ldots,J\}).
        \end{array}\right.
\end{align}
Similarly, for step \textbf{(R)}, we unfold by replacing the adjoint of the regularization operator $\Bb_1$ with $\tilde{\Bb}_1 = \begin{bmatrix} \tilde{\nabla}_1^\top  & \cdots & \tilde{\nabla}_j^\top \end{bmatrix}^\top$.
Setting, for every $j\in \{1,\ldots,K\}$, $\nu_{n,j} = \gamma_n\tau_j^{-1}$
yield the following regularization layer $\mathcal{L}_{\rm R}$: 
\begin{align}
&\mbox{\textbf{Regularization layer ($\mathcal{L}_{\rm R}$)
:}}\nonumber\\
&\quad
\left\lfloor
\begin{array}{l}
\boldsymbol{x}_n = \proj_{[0,+\infty[^{L}}(\wb_n)\\
\mbox{For $j \in \{1,\ldots, J\}$} \\
\quad
\left\lfloor
\begin{array}{l}
(\forall \ell \in \{1,\ldots, L\}) \\
 \quad \displaystyle (\zb^j_{n+1})_{\ell}= \frac{\left(\zb^j_n + \nu_{n,j} \nabla_j \boldsymbol{x}_n\right)_{\ell}}{\max\big\{1,\|\left( \zb^j_n + \nu_{n,j} \nabla_j \boldsymbol{x}_n\right)_{\ell}\|_2/ \alpha_{j,\ell}\big\}} \end{array}\right.\\
\wb_{n+1} = \wb_n - \Mb^{-1} \sum_{j=1}^J \tilde{\nabla}_j (\zb^j_{n+1} -\zb^j_n)\\
\zb_{n+1}^0 = \zb_n^0.
\end{array}\right.
 \label{eq:layerR}
\end{align}
Note that $\mathcal{L}_{R}$ does not involve $\overline{\xb}$.

\subsection{Total architecture}

\begin{figure*}[htbp]
     \centering
     \begin{subfigure}[b]{\textwidth}
         \centering
         \includegraphics[width=0.6\textwidth]{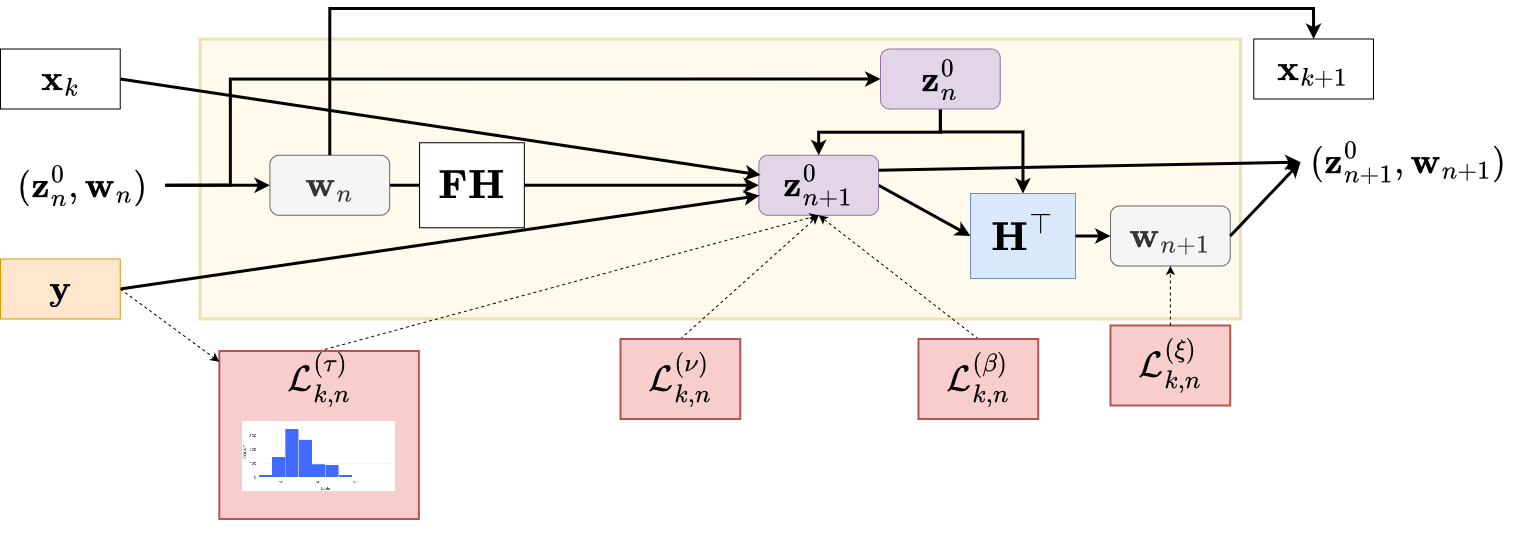} 
\caption{Schematic view of a $N_{k}$-th layer $\mathcal{L}_{D}$ \eqref{eq:layerD}. The layer relies on $\overline{\xb} = \xb_k$, the $k$-th reweighted iterate to generate the next reweighted iterate $\xb_{k+1}$. The layer takes as inputs $\wb_{n}$, $\zb_{n}^0$ from the previous layers. The projections $\yb$ are also used as input.}
\label{fig:U_Da}
     \end{subfigure}
     
     \begin{subfigure}[b]{\textwidth}
         \centering
         \includegraphics[width=0.7\textwidth]{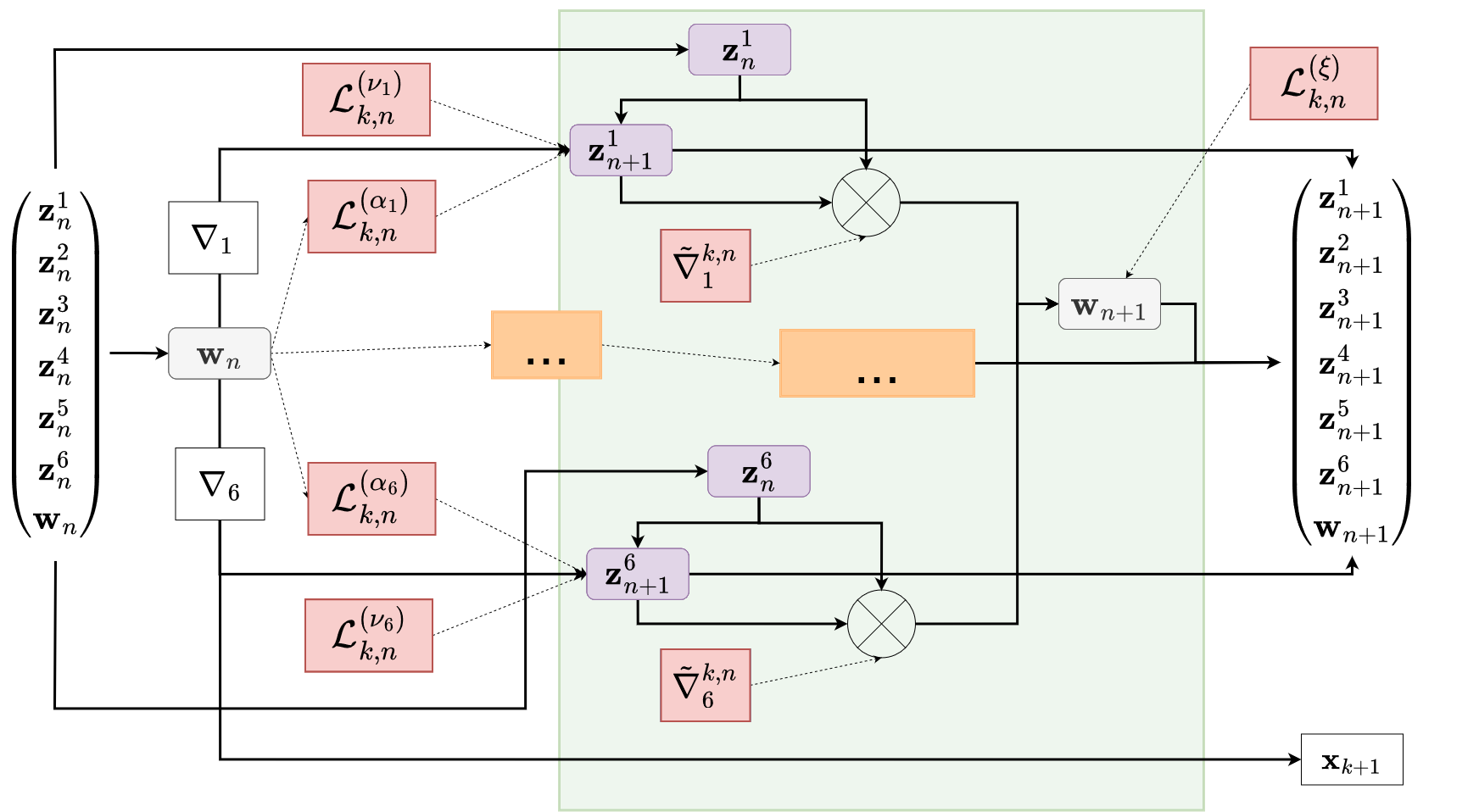} 
\caption{Schematic view of a $N_{k}$-th layer $\mathcal{L}_{R}$ \eqref{eq:layerR}. The layer takes as inputs $\wb_{n}$, $(\zb_{n}^j)_1^J$ from the previous layer and generates the next reweighted iterate $\xb_{k+1}$. The update of parameters for $j\in \{2,\ldots, 5\}$ is hidden in the orange block for the sake of readability. Only parameters $\alpha_j$ depend on the input.}
\label{fig:U_Ra}
     \end{subfigure}
     
\begin{subfigure}[b]{1\textwidth}
         \centering
         \includegraphics[width=18cm]{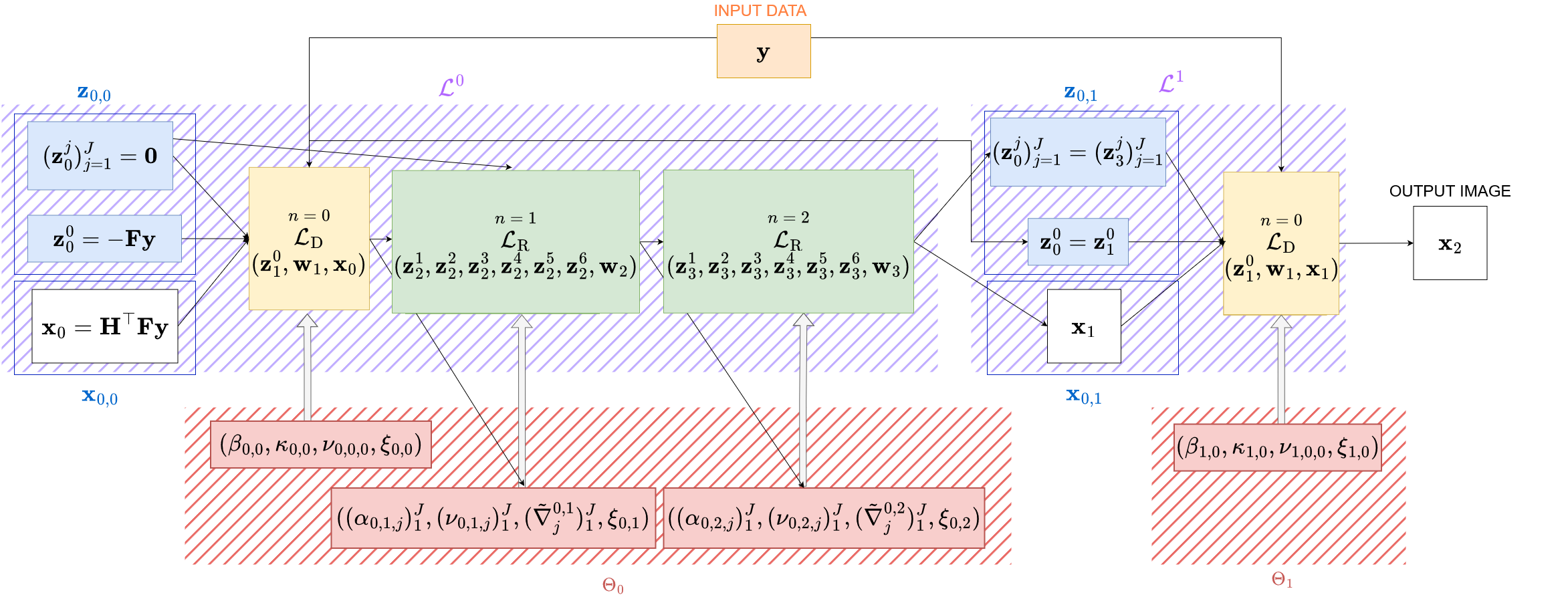} 
\caption{U-RDBFB in the case where $\mathcal{A} = \mathcal{L}^1 \circ \mathcal{L}^0_2$ where $\mathcal{L}^0=  (\mathcal{L}_{R})^2 \circ \mathcal{L}_{D}$ and $\mathcal{L}^1_2=\mathcal{L}_{D}$ (i.e., $K=2$, $N_0 = 3$, $N_1 = 1$). Red blocks represent the hidden structures to infer all the parameters $\theta \in \Theta$. 
When $k=1$, the dual variables of DBFB are initialized with the values of the dual variables at the end of the previous $N_0$ iterations of DBFB ($k=0$).
} 
\label{fig:U_DBFB}
     \end{subfigure}
     \caption{Architecture of U-RDBFB}
 \end{figure*}

The total architecture of U-RDBFB, denoted $\mathcal{A}$, can be summarized as 
\begin{align}\label{def:A}
    \mathcal{A} = \mathcal{L}^{K-1} \circ \cdots \circ \mathcal{L}^0. 
\end{align}
For $ 0\leq k \leq K-1$, $\mathcal{L}^{k}$ corresponds to a sequence of $N_k$ layers $\mathcal{L}_D$ or $\mathcal{L}_R$ and implements the following update:
    \begin{equation}
       (\zb_{0,k+1}, \xb_{k+1}) = \mathcal{L}^{k}(\zb_{0,k}, \xb_k;\Theta_{k}),
        \end{equation}
where 
\begin{itemize}
    \item $\xb_k$ is the current reweighted estimate ${\overline{\xb}}$ in $\mathcal{L}_{\rm R}$-$\mathcal{L}_{\rm D}$ ($\xb_0 = \Hb^\top \Fb \yb$).
    \item $\xb_{k+1}$ is the next reweighted estimate; it is equal to $\boldsymbol{x}_{N_k-1}$ given by the $N_k$-th layer $\mathcal{L}_{\rm R}$-$\mathcal{L}_{\rm D}$. 
    \item $\zb_{0,k} \in \RR^T\times (\RR^{2L})^J$ is the initial value of the variables $(\zb_{0}^j)_{0\le j \le J}$ for layers $\mathcal{L}_{\rm R}$-$\mathcal{L}_{\rm D}$ (for $k=0$, $(\zb^j_0)_{j=1}^J$ are initialized to zero while $\zb^0_0$ is set to $-\Fb \yb$).
    \item $\zb_{0,k+1} \in \RR^T\times (\RR^{2L})^J$ is equal to $(\zb_{N_{k-1}}^j)_{0\le j \le J}$ given by the $N_k$-th layer $\mathcal{L}_{\rm R}$-$\mathcal{L}_{\rm D}$
    \item $\Theta_{k}$ is the vector of trainable parameters. The parameters are layer-dependent, so we index them by $k$ and $n$.\\
For layer $\mathcal{L}_{\rm D}$, the parameters are those of the Cauchy function $(\beta_{k,n},\kappa_{k,n})$, the one of the quadratic regularization $\xi_{k,n}$, and a single step size $\nu_{k,n,0}$. \\
For layer $\mathcal{L}_{\rm R}$, the regularization parameters $(\alpha_{k,n,j,\ell})_{1\le j\le J, 1\le \ell \le L}$, $\kappa_{k,n}$, $\xi_{k,n}$, and step sizes $(\nu_{k,n,j})_{j=1}^J$ are learned as well as the surrogates {$(\tilde{\nabla}^{k,n}_j)_{j=1}^J$} to the adjoints of operators $(\nabla_j^\top)_{j=1}^J$. \\
To infer all these parameters, we introduce learning modules in each layer $(\mathcal{L}_{k,n}^{(\theta)})_{n=0}^{N_k-1}$ for $\theta \in \Theta_{k}$ (implementation discussed in Appendix~\ref{a:PL}).
\end{itemize}
Schematic views of layers $\mathcal{L}_{\rm D}$ and $\mathcal{L}_{\rm R}$ can be found in Figure~\ref{fig:U_Da} and Figure~\ref{fig:U_Ra}, and a composition $\mathcal{A}$ of such layers is displayed in Figure~\ref{fig:U_DBFB}.\\
\\
Here we propose using $K=7$ in \eqref{def:A} with $N_k=4$ for each $k \in \{0, \cdots, K-1\}$, 
resulting in a total of 28 layers:
\begin{align}\label{eq:URDBFB_dev}
    \mathcal{L}^0 = \cdots = \mathcal{L}^6 =  (\mathcal{L}_{R} \circ \mathcal{L}_{D})^2.
\end{align}

\subsection{Incremental training strategy}

For each learning module, we circumvent the issue of optimizing the initial values of their parameters using an incremental training strategy, as sometimes advocated for when initializing the weights of recurrent neural networks \cite{safran2016quality,martens2011learning}.\\ The learning in each layer $(k,n)$ ($n\geq1$) starts by considering all the previous layers from $(0,0)$ to $(k,n-1)$ with their past trained parameters as an initialization. This means that an increasing number of layers are trained simultaneously. In the last step, all layers are trained end-to-end.  

\section{Experiments}\label{sec:exp}

We now describe our experimental setup. 
First, we illustrate the benefits of using a Cauchy-based data fidelity function. We compare the results of our RDBFB iterative approach to a simpler DBFB scheme minimizing the same cost function but with $g$ in \eqref{eq:pb0} replaced with the $\ell_2$ norm.
\\
Second, we comment on the improved performance brought by our unfolding strategy (learning of adjoints and parameters as well as the use of the ramp filter), and so U-RDBFB (Section~\ref{sec:unfold}) is compared to the original RDBFB algorithm.
\\ 
Third, we compare U-RDBFB to several state-of-the-art reconstruction methods and comment on the transfer of performance over different synthetic datasets.

\subsection{Datasets}

We used three datasets for evaluation. 

\subsubsection{Abdomen dataset}
Our first dataset consists of 2D images obtained from {60 CT volumes} of size $512\times 512 \times 512$ from the lower lungs to the lower abdomen of 60 patients, which were extracted from the public dataset \textit{CT Lymph Nodes} from \href{https://www.cancerimagingarchive.net/}{https://www.cancerimagingarchive.net/}. These volumes correspond to fully sampled CT reconstructions. They were made isotropic by interpolating the axial slices. A total of 50 out of 512 slices were kept per volume. 
We randomly added intense metallic wires between 3000 and 5000 Hounsfield units (HU) of varying sizes on the axial slices. We shifted the HU values of the images by 1000 so that air is 0 HU and water is 1000 HU, using 
$\boldsymbol{a}_{\rm tissue} \mapsto
(\boldsymbol{a}_{\rm tissue} - \mu_{\rm water})\times (1000/\mu_{\rm water})$,
where $\mu_{\rm water}$ is the value of the attenuation coefficient of water equal to 0.017 mm$^{-1}$ and $\boldsymbol{a}_{\rm tissue} \in \RR^Q$ is the initial vector of attenuation values.
Finally the $512 \times 512$ slices $\overline{\xb}_{\rm P}$ were normalized between $[0,1]$ (a value of $1$ corresponding to an object of HU intensity equal to $5000$).\\
To eliminate bias with respect to model discretization, projections were simulated for each slice of each volume in a 2D parallel geometry with a short detector of {600 bins (bin size equal to half a pixel size, i.e., 0.5 mm) and an angular density of 110 projections over $180^{\degree}$ through to the forward projector $\Hb_{\rm P} \in \RR^{512^2\times (110\times 600)}$. The projections were then rebinned by a factor 2 (operator $\Rb \in \RR^{300 \times 600}$)}.
Noisy projections $\yb = (y_t)_{t=1}^T$ are computed as
\begin{equation*}\label{eq:BL_1E}
    (\forall t \in \{1,\ldots,T\}) \quad y_t = \mu \log \left(\frac{I_0}{\mathcal{P}(I_0 \exp{(- \mu ({\Rb} \Hb_{\rm P} \overline{\xb}_{\rm P})_t)})}\right),
\end{equation*}
where we set $\mu = \mu_{\rm water}/1000$, $I_0 = 10^4$, and, for some $\delta>0$, $\mathcal{P}(\delta)$ denotes a realization of a Poisson law with mean $\delta$.  
In this context, the ROI was a centered disk of diameter $300$. 
The resulting pairs of axial slice/projections $(\overline{\xb}_{\rm P},\yb)$ were split into a training of 2500 pairs from a pool of 50 patients and a testing set of 500 pairs from 10 other patients.

\subsubsection{Head dataset}
We used a second dataset containing 2D images extracted from {10} CT high-dose brain reconstructions. These volumes are from the public repository \textit{2016 Low Dose CT Grand Challenge} from \href{https://www.cancerimagingarchive.net/}{https://www.cancerimagingarchive.net/}. {After extracting 50 slices of size $512\times512$ per volume (pixel size of 0.5 mm), we performed the same processing as for the Abdomen dataset (addition of intense wires, normalization, projection, rebinning) for generating a testing set of 500 pairs of axial slices/projections $(\overline{\xb}_{\rm P},\yb)$.}

\subsubsection{Geometrical dataset}
Our third dataset was created using the toolbox TomoPhantom \cite{tomophantom}. 500 geometrical 2D piecewise-constant phantoms were randomly generated on a $512\times512$ grid and normalized between 0 and 1. Again, we performed the same processing as for the Abdomen dataset for generating a testing set of 500 pairs of axial slices/projections $(\overline{\xb}_{\rm P},\yb)$. 

\subsection{Training details for U-RDBFB}

Let $i\in\{1, \ldots,I\}$ be the index covering all $I=2500$ instances of the training set. The reconstruction grid ($G$) is a disk of diameter $400$.
Let $\xb^*_{\rm G, i} \in \RR^L$ be the output of U-RDBFB for a given projection input $\yb_i$. Our network is thus designed to minimize $\sum_{i=1}^I \ell(\Cb_{\rm G} \xb^*_{{\rm G},i}, \Cb_{\rm P} \overline{\xb}_{{\rm P},i})$, where $\Cb_{\rm G}$ is a cropping operator which extracts the ROI from the grid $G$, $\Cb_{\rm P}$ is a cropping operator which extracts the ROI from the entire $512\times 512$ grid, and $\ell$ is the loss retained for training the network. For all instances of the training set, 
\begin{equation}\label{eq:loss}
    \ell(\Cb_{\rm G} \cdot, \Cb_{\rm P} \overline{\xb}_{\rm P,i}) = \frac{1}{I} \|\Cb_{\rm G} \cdot- \Cb_{\rm P} \overline{\xb}_{\rm P,i}\|^2,
\end{equation}
corresponding to the MSE loss. 
\\
We implemented U-RDBFB following \eqref{eq:URDBFB_dev} in Pytorch, using a Tesla V100 32 Gb GPU. We used six epochs for training each layer $\mathcal{L}_{\rm R}$ and ten epochs for training each layer $\mathcal{L}_{\rm D}$; the only exception was the last layer, for which we used 20 epochs. The learning rate is decreased with a step decay by a factor of 0.99 from $10^{-2}$ every four epochs. The batch size for each epoch varied from 20 to 8 as the number of trained layers increased. 
We employed the toolbox TorchRadon \cite{Ronchetti2020} to include Pytorch-compatible parallel-beam tomographic operators in all architectures. Standard auto-differentiation tools can compute all necessary derivatives for backpropagation. The training procedure takes about one day and a half.

\subsection{Competing methods}
The quantitative metric used to assess the reconstruction quality of $\Cb_{\rm G} \xb^*_{\rm G, i}$ is the PSNR.
We also evaluate the reconstruction performance using the structural similarity index (SSIM), the PieApp value \cite{prashnani2018pieapp}, and the Mean Absolute Error (MAE) of the difference between $\Cb_{\rm G} \xb^*_{\rm G, i}$ and $ \Cb_{\rm P} \overline{\xb}_{\rm P,i}$. \\
We compare U-RDBFB with FBP, an iterative method, and four deep-learning methods that we describe hereinafter.

\subsubsection{FBP}
This analytical method consists of computing $\Hb_{\rm ROI}^\top \Fb \yb$, where $\Hb_{\rm ROI} \in \RR^{300^2 \times (110\times 300)}$.
As is commonly the case when applying FBP on truncated data, we extrapolated the projections prior to ramp filtering (using anti-symmetric padding).

\subsubsection{RDBFB algorithm}
For completeness, we perform comparisons with the iterative method proposed in Section~\ref{sec:theory}. For each reweighted iteration $k$, we used $N_k = 10$ DBFB iterations alternating between data and regularization steps (1:1 correspondence). For an easier manual tuning of the hyperparameters, instead of using $J=6$ as in U-RDBFB, we set $J=1$ so that STV reduces to TV and $\alpha_{1, \ell} \equiv \alpha_1$, for all $\ell \in \{1, \cdots, L\}$. The remaining cost function parameters ($\xi$, $\kappa$, $\beta$) are selected by optimizing PSNR on the training set via a grid search.

\subsubsection{Post-processing U-net}
The third competing method is the CNN proposed in \cite{Jin2017,Han2016}, which is a post-processing of FBP. It relies on a trained residual U-net, with a depth of 4 levels, filters of size 32, and batch normalization to improve the stability of training. 

\subsubsection{Preconditioned Neumann Network (PNN)}
\begin{figure}[h]
\centering
\includegraphics[width=8.5cm]{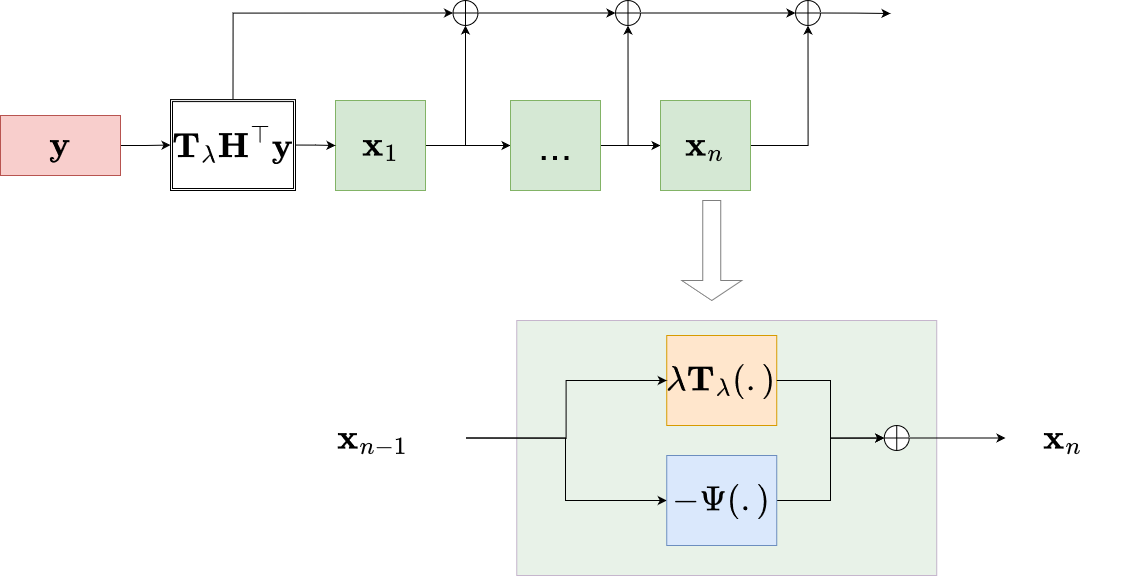} 
\caption{Architecture of PNN \cite{Gilton2020NeumannNF}: The network maps a linear function of the measurements $\Tb_\lambda \Hb^\top \yb$ to a reconstruction $\xb_n$ by successive applications of an operator of the form $\lambda \Tb_\lambda -  \Psi$, while summing the intermediate outputs of each block. All instances of $\Tb_\lambda$ are replaced by an unrolling of 10 iterations of the conjugate gradient algorithm. $\Psi$ is a trained network and the scale parameter $\lambda$ is also trained.}
\label{fig:Neumann}
     \end{figure}

Our fourth competing method is a preconditioned Neumann network (PNN) initially introduced in \cite{Gilton2020NeumannNF} for MRI reconstruction. It builds on a method for solving Problem \eqref{eq: generic_cost_func} with $f(\xb)=\frac12 \|\Hb \xb - \yb\|^2$. For a differentiable function $r$, the resulting minimizer reads 
\begin{equation}
(\Hb^\top \Hb + \nabla r) {\xb} = \Hb^\top \yb,
\end{equation}
which can be rewritten as
\begin{equation}
    (\Hb^\top \Hb +\lambda \Id){\xb} + ( \nabla r - \lambda \Id) {\xb} = \Hb^\top \yb.
\end{equation}
Setting $\Tb_{\lambda} = (\Hb^\top \Hb + \lambda \Id)^{-1}$ yields 
\begin{equation}
    (\Id - \lambda \Tb_{\lambda} + \Tb_{\lambda} \nabla r) {\xb} = \Tb_{\lambda} \Hb^\top \yb.
\end{equation}
Using the Neumann identity
    $\Bb^{-1} =  \sum_{n=0}^{\infty} (\Id -  \Bb)^n$,
the authors derive the architecture of PNN with $N \in \NN^*$ layers (see Figure \ref{fig:Neumann})
\begin{equation}
   (\lambda \Tb_{\lambda} - \Tb_{\lambda}  \nabla r)^N \circ \Tb_{\lambda} (\Hb^\top \yb).
\end{equation}
All instances of $\Tb_{\lambda}$ are applied approximately using an unrolling of 10 iterations of the conjugate gradient algorithm. \\
The operator $\Tb_{\lambda} \nabla r$ is replaced by a U-net, denoted by $\Psi$, which has the same architecture as the aforementioned U-net without the residual connection. The weights of the U-net are shared for all layers. Following \cite{Gilton2020NeumannNF}, no batch normalization is used. The inner U-net has a depth of 4, the learning rate is set to $10^{-4}$, and the initial value for $\lambda$ is $0.01$. 
We choose $N=3$. One feature of PNN compared to other deep unfolding networks is that it contains skip connections. 

\subsubsection{ISTA-net}
Our fifth competing method is ISTA-net, derived from the work of \cite{Zhang2018}. ISTA-net is designed to solve Problem \eqref{eq: generic_cost_func} for $f(\xb)=\frac12 \|\Hb \xb - \yb\|^2$, and $r(\xb) = \lambda \|\Wb \xb\|_1$ ($\lambda>0$), where operator $\Wb$ is not known a priori but learned.
$\Wb$ is an orthogonal linear operator in the initial ISTA algorithm, whose iteration reads 
\begin{equation}
    \xb_{n+1} = \Wb^\top \operatorname{soft}\left( \Wb(\xb_n - \tau \Hb^\top (\Hb \xb_n - \yb),\lambda  \tau \right),
\end{equation}
where $\operatorname{soft}$ is the soft-thresholding operation and $\tau>0$ is the gradient step size.
In ISTA-net, the authors replace $\Wb$ and $\Wb^\top$ by two decoupled nonlinear operators namely $\Ab_n \circ \operatorname{ReLU} \circ \Bb_n$ and $\Cb_n \circ \operatorname{ReLU} \circ \Db_n$ (see Figure \ref{fig:PGAnetPlus}). The property of orthogonality of $\Wb$ is not imposed but favored during training by adding a term, weighted by $\chi \in \RPP$, penalizing the difference between $(\Cb_n \circ \operatorname{ReLU} \circ \Db_n) \circ (\Ab_n \circ \operatorname{ReLU} \circ \Bb_n) \xb_n$ and $\xb_n$ in the loss function. Each $\Ab_n$, $\Bb_n$, $\Cb_n$ and $\Db_n$ is a 2D convolutional operator. $\Bb_n$ and $\Cb_n$ are associated with a kernel of size $3\times3$ and {32} input and output channels; $\Ab_n$ has 1 input channel and {32} output channels and vice-versa for $\Db_n$. As suggested by the authors, we learn these convolutional operators as well as $\lambda$ and $\tau$, which are allowed to vary at each iteration. \\
Experiments are carried out with 10 layers, $\chi = 0.1$, $\xb_{0}$ is the FBP reconstruction, $\lambda$ and $\tau$ are initialized to $100$ and $0.1$ respectively.

\subsubsection{PD-net}
\begin{figure*}[h!]
     \centering
     \begin{subfigure}[b]{\textwidth}
\centering
\includegraphics[width=11cm]{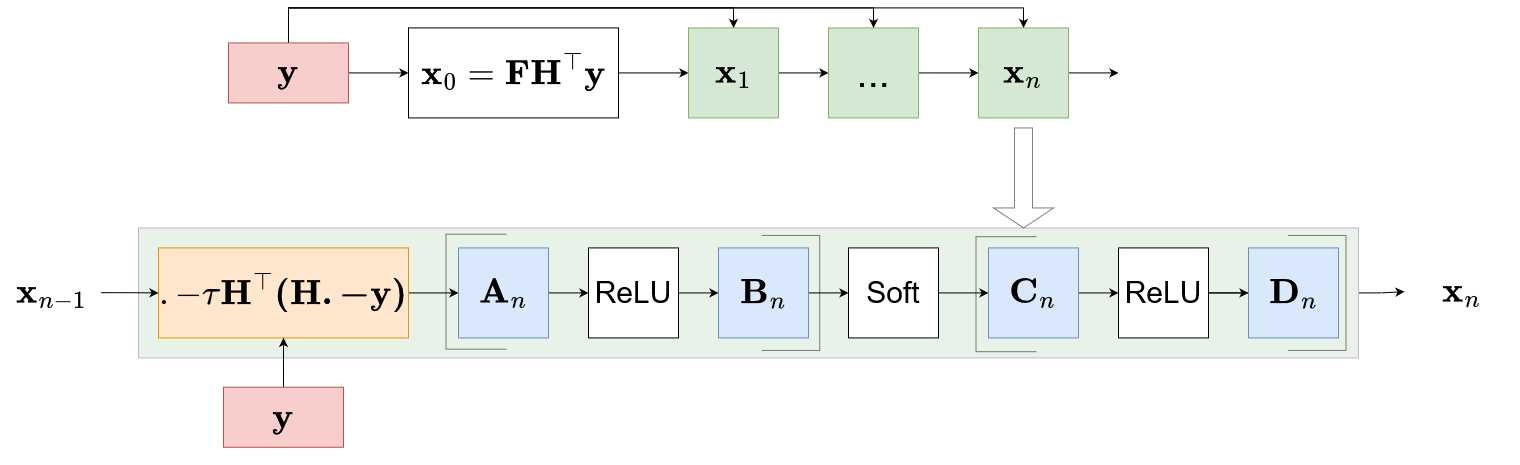} 
\caption{Architecture of ISTA-net \cite{Zhang2018}: Each layer is composed of a gradient step followed by the application of a nonlinear operator, which is the combination of two learnable linear convolutional operators ($\Ab_n$, $\Bb_n$) separated by a ReLU, a soft-thresholding operation and then two other learnable linear convolutional operators ($\Cb_n$, $\Db_n$) separated by a ReLU. The property $(\Cb_n \circ \operatorname{ReLU} \circ \Db_n) \circ (\Ab_n \circ \operatorname{ReLU} \circ \Bb_n) = \Id$ is favored during training.}
\label{fig:PGAnetPlus}
     \end{subfigure}
     
\begin{subfigure}[b]{\textwidth}
\centering
\includegraphics[width=12cm]{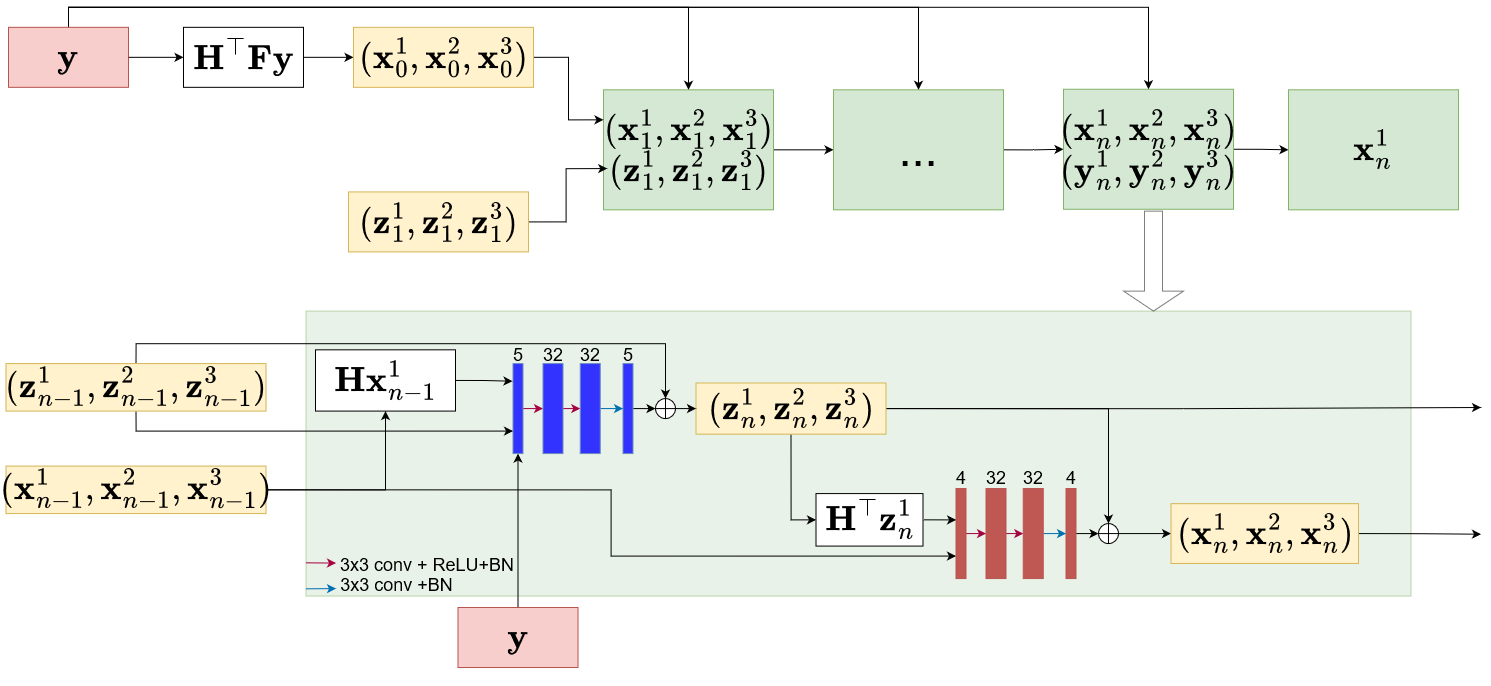} 
\caption{Architecture of PD-net \cite{Adler2018}: The red and blue boxes represent the primal and dual networks, respectively. Buffers of 3 primal $(\xb_{n}^1,\xb_{n}^2,\xb_{n}^3)$ and dual $(\zb_{n}^1,\zb_{n}^2,\zb_{n}^3)$ estimates are used at each iteration. The initial primal estimates are set to the FBP reconstruction given by $\Hb^\top \Fb \yb$, and the initial dual estimates are set to zero.}
\label{fig:PDnet}
     \end{subfigure}
     \caption{Two competing unfolded proximal algorithms}
 \end{figure*}

The last competing method is the learned Primal-Dual (PD-net) introduced in \cite{Adler2018} by unrolling the Primal-Dual Hybrid Gradient (PDHG) optimization algorithm\cite{Cheng2019}. The authors consider Problem \eqref{eq: generic_cost_func} with a more generic data fidelity term $f(\xb) = G(\Hb \xb; \yb)$. They replace both the proximity operators of $G$ and $r$ in PDHG by residual CNN so that one layer $n$ of their network reads
\begin{align}
    \zb_{n+1} &= \operatorname{CNN}(\zb_n+\sigma \Hb \tilde{\xb}_n; \yb)\\
    \xb_{n+1} &= \operatorname{CNN}(\xb_n - \tau \Hb^\top \zb_{n+1})\\
    \tilde{\xb}_{n+1} &= \xb_{n+1} + \gamma(\xb_{n+1}-\xb_{n}).
\end{align}
The CNNs act both in the image and projection domains. Furthermore, buffers of previous iterates of size $N_p \in \NN$ in the primal domain (image) and of size $N_d \in \NN$ in the dual domain (projection) are kept to enable the network to learn an acceleration. 
We used {$9$} layers, $N_d = N_p = 3$, and 32 filters in the convolutional layers. This network is illustrated in Figure~\ref{fig:PDnet}. 

\begin{table*}[b]
\centering
\begin{tabular}{|c|c|c|c|c|c|}
\hline
 & U-RDBFB  & U-net & PNN  & PD-net & ISTA-net  \\
\hline 
$|\Theta|$ & $2.3169 \times 10^4$ & $1.9278\times 10^6$   &$1.9278\times 10^6$  &$2.5470\times 10^5$  &$1.7109\times 10^5$ \\ \hline
\end{tabular}
\caption{Number of learnable parameters ($\Theta$)}
\label{tab:capacity}
\end{table*}

The competing networks were also trained with the MSE loss (using \eqref{eq:loss} for unfolding networks and a regularization term for ISTA-net weighted by $\chi$) in a standard end-to-end manner. {The number of epochs was chosen such that the training and testing losses have stabilized.}
Note that codes are publicly available for these networks. We re-implemented them in Pytorch and kept the setting of the parameters advocated by the authors, except for PNN, for which we reduced the number of layers to 3 to obtain a stable behavior for training. The total number of parameters of each network is reported in Table~\ref{tab:capacity}.

\begin{figure*}[h!]
 \centering
  \begin{subfigure}[b]{0.2\textwidth}
     \centering
         \includegraphics[width=\textwidth]{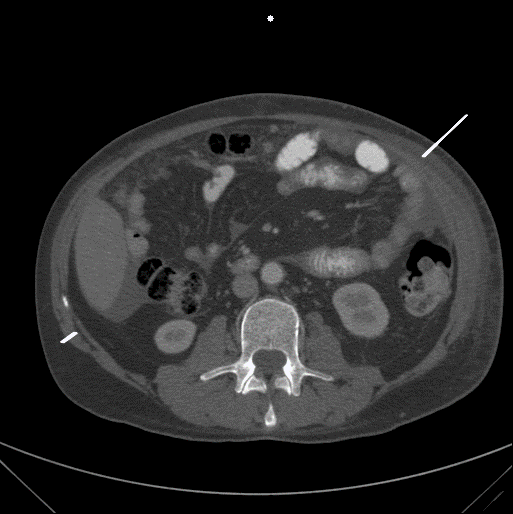}
     \caption{}
    \label{fig:GT}  
 \end{subfigure}
 \begin{subfigure}[b]{0.3\textwidth}
     \centering
 \includegraphics[width=\textwidth]{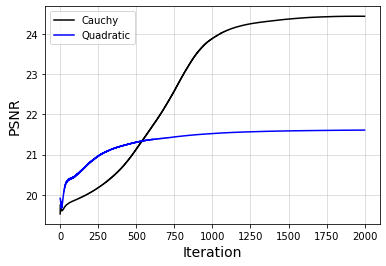}
     \caption{}
    \label{fig:PSNR_c_Q}
 \end{subfigure}
      \begin{subfigure}[b]{0.35\textwidth}
     \centering
     \includegraphics[width=\textwidth]{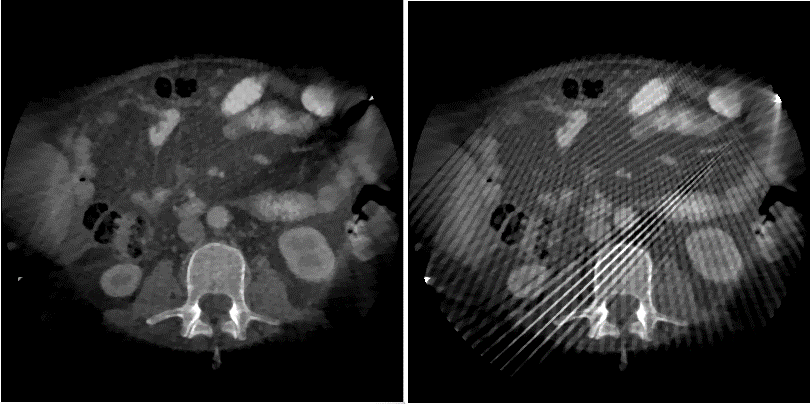}
     \caption{}
    \label{fig:img_C_Q}
 \end{subfigure}
 \\
 \begin{subfigure}[b]{\textwidth}
     \centering
\includegraphics[width=0.8\textwidth]{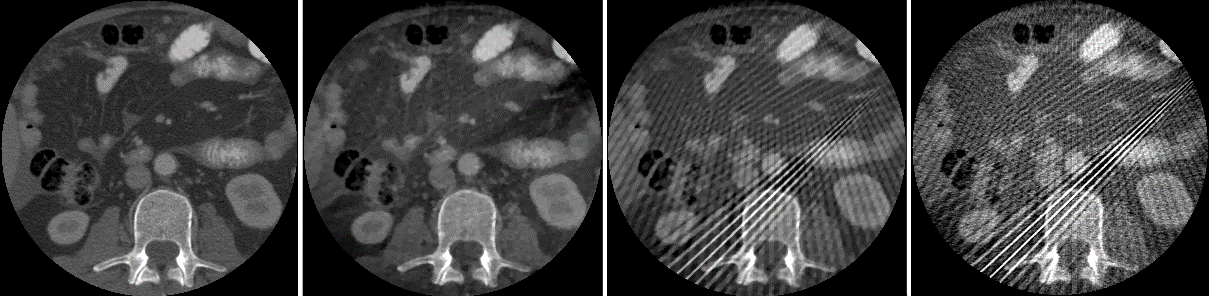}
\caption{}
\label{fig:ROIs_iter}      
\end{subfigure}
\\
 \begin{subfigure}[b]{\textwidth}\centering
\includegraphics[width=0.8\textwidth]{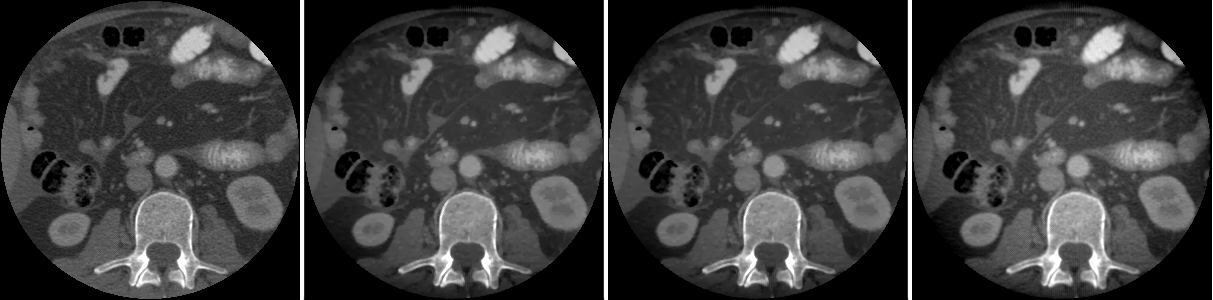}
\caption{}
\label{fig:ROIs_iter2}
\end{subfigure}
\caption{(a) Ground truth $\overline{\xb}_{\rm P}$. (b) Evolution of the PSNR along iterations using a Cauchy or quadratic data term for 110 projections. (c) Reconstructed extended ROIs using 110 projections, a Cauchy data fidelity, and a quadratic data fidelity. (d) Reconstructed ROIs using 110 projections. (e) Reconstructed ROIs using 600 projections. From left to right: Ground truth, reweighted DBFB with Cauchy fidelity, DBFB with quadratic fidelity, FBP.}
\end{figure*}

\section{Results}\label{sec:res}

\subsection{Assessing the benefits of the Cauchy fidelity term}

Figure \ref{fig:img_C_Q} shows the full reconstructed images on grid $G$ of size $400\times 400$ obtained using the DBFB algorithm with a quadratic fidelity term and the reweighed DBFB algorithm with a Cauchy fidelity on a test instance of the Abdomen dataset (shown in Figure \ref{fig:GT}). Since the two data fidelity terms can be put into our optimization framework Alg. \ref{alg:rewighted}-Alg. \ref{alg:DBFB}, the comparison is straightforward.
Figure \ref{fig:ROIs_iter} shows the corresponding ROIs as well as the FBP reconstruction. The full image contains two intense objects out of the ROI and at the border of the reconstruction grid. In the solution obtained using the quadratic data fidelity term, the reduction of sub-sampling streaks is selective; only the streaks originating from objects within $G$ have been eliminated in the ROI. When trading the quadratic term with a Cauchy term, as we proposed, the intensity of these streaks is reduced. This artifact reduction translates into an improvement of the PSNR as shown in Figure \ref{fig:PSNR_c_Q}. 
Figure \ref{fig:ROIs_iter2} shows the ROIs obtained using the same reconstruction methods and grid size when increasing the number of projections from 110 to 600. The images now look identical and close to the ground truth. This observation highlights that, for relatively 'clean' data (no modeling of beam hardening and scattering), the benefits of using a Cauchy fidelity over a quadratic fidelity emerge when data is sub-sampled. \\
We have shown that by using a regularized cost function with a Cauchy fidelity term and, thus, a more complex optimization framework, we can successfully reconstruct truncated data on a short grid and that the reconstruction is at least as good as the one obtained with quadratic fidelity and even better when the data are sub-sampled.

\begin{figure*}[h!]
 \centering
 \begin{subfigure}[b]{0.4\textwidth}
     \centering
 \includegraphics[width=0.8\textwidth]{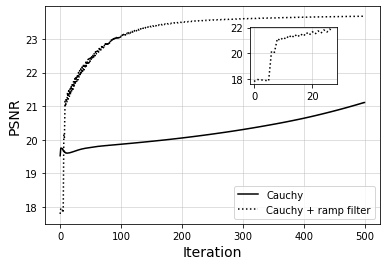}
      \caption{}
\label{fig:PSNR_Ramp}
 \end{subfigure}
      \begin{subfigure}[b]{0.4\textwidth}
     \centering
     \includegraphics[width=0.8\textwidth]{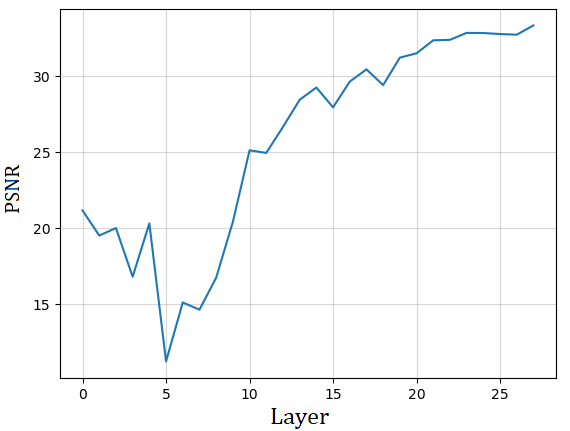}
      \caption{}
\label{fig:PSNR_URDBFB} 
 \end{subfigure}
 \\
 \begin{subfigure}[b]{\textwidth}
     \centering
\includegraphics[width=0.8\textwidth]{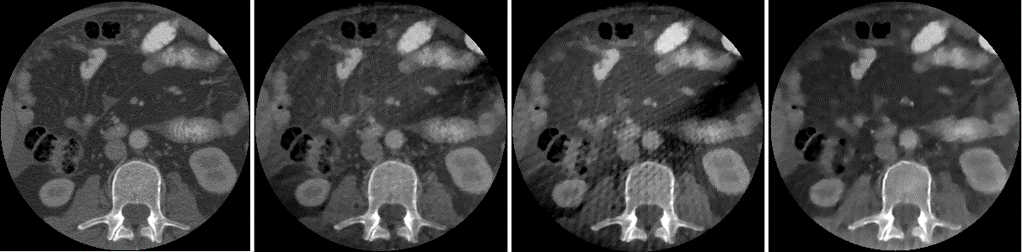}
\caption{}
\label{fig:ROIs_iter_udbfb}      
\end{subfigure}
\caption{(a) Evolution of the PSNR along iterations using a Cauchy fidelity term with and without the ramp filter for the example of Fig \ref{fig:GT}. (b) Evolution of the PSNR along layers in U-RDBFB for the example of Fig \ref{fig:GT}. (c) Reconstructed ROIs. From left to right: Ground truth, Cauchy with ramp filter (500 iterations of the modified reweighted DBFB), Cauchy with ramp filter (28 iterations of the modified reweighted DBFB), U-RDBFB.}
\end{figure*}

\subsection{Comparing iterative RDBFB algorithm with U-RDBFB}
Figure \ref{fig:PSNR_Ramp} shows the evolution of the PSNR along the iterations when inserting the ramp filter in the reweighted DBFB algorithm, re-tuning the regularization strength, and still using the same data. The PSNR stagnates around 300 iterations with the ramp filter while it stagnates around 1250 iterations without it (see Figure \ref{fig:PSNR_c_Q}). Thus applying the ramp filter on the reprojection error before backprojection can empirically accelerate convergence without degrading the solution (reconstructed ROI displayed in Figure \ref{fig:ROIs_iter_udbfb}) in an early stopping scenario. This motivates our translation of a data iteration of DBFB to a data layer of U-RDBFB which embeds the ramp filter. It also provides empirical evidence that performance can be optimized by introducing mismatched adjoints without learning.\\
U-RDBFB also includes learned parameters, especially adjoints to the STV operators. It performs a total of 28 RDBFB iterations. Figure \ref{fig:ROIs_iter_udbfb} compares the reconstruction ROI obtained with U-RDBFB, 500 and 28 iterations of the reweighed DBFB algorithm with the ramp filter. We see that after 28 iterations of the reweighed DBFB algorithm, there is a local offset near the intense object, and some streaks remain. On the contrary, the image obtained with U-RDBFB does not contain these artifacts. It is similar to the ground truth and slightly smoother than the reconstruction after 500 iterations of the reweighed DBFB algorithm but with approximately 18 times fewer iterations.

\subsection{Comparing U-RDBFB with deep learning methods on the Abdomen dataset}

\begin{figure}[h!]
 \centering
 \begin{subfigure}[b]{0.45\textwidth}
     \centering
 \includegraphics[width=\textwidth]{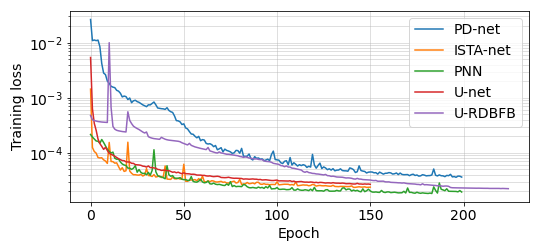}
     \caption{}
    \label{fig:loss1}
 \end{subfigure}
 \\
      \begin{subfigure}[b]{0.45\textwidth}
     \centering
     \includegraphics[width=\textwidth]{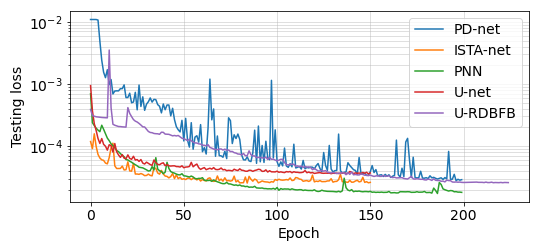} 
     \caption{}
    \label{fig:loss2}
 \end{subfigure}
\caption{{(a) MSE on the training set as a function of the epoch number. (b) MSE on the testing set as a function of the epoch number.}}
\end{figure}

\begin{sidewaysfigure*}[htbp]
\centering
 \begin{subfigure}[b]{0.7\textwidth}
    \centering
         \includegraphics[width=\textwidth]{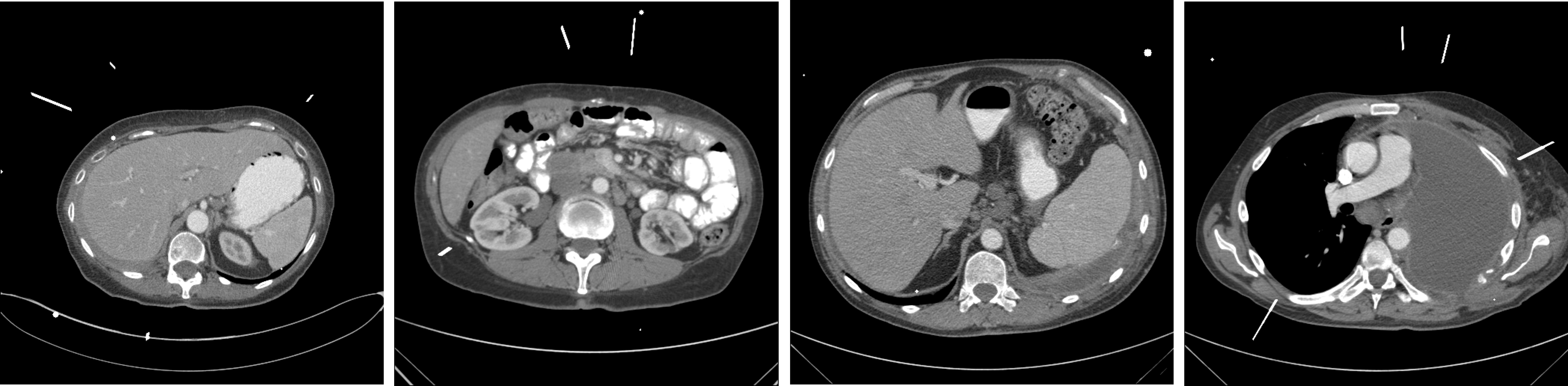}
    \caption{Four ground truths of the testing set of the Abdomen dataset}
    \label{fig:GT_1}
    \end{subfigure}
    
     \begin{subfigure}[b]{0.6\textwidth}
    \centering
         \includegraphics[width=0.9\textwidth]{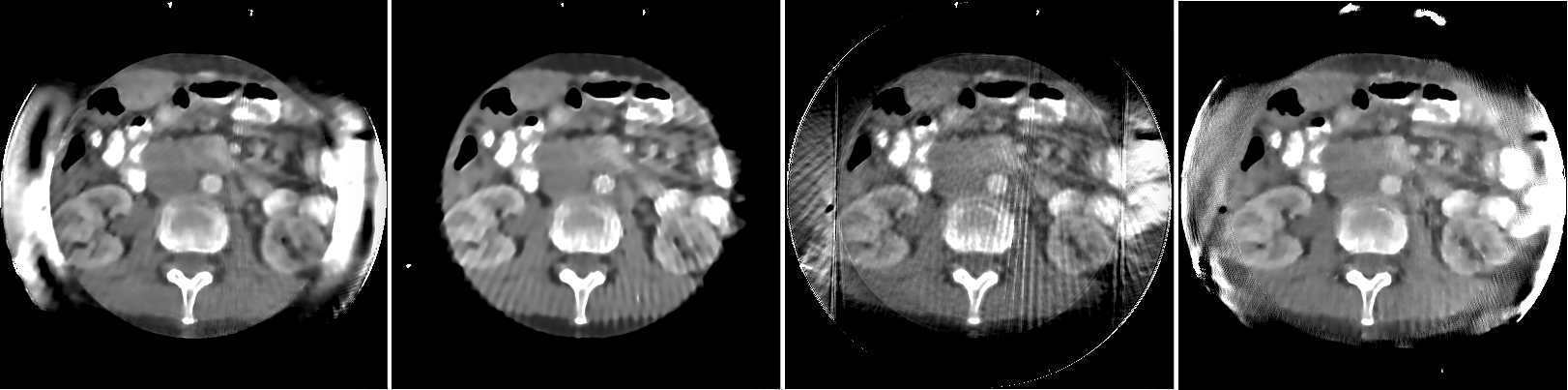}
    \caption{Reconstructed images on grid $G$ using PNN, ISTA-net, PD-net, and U-RDBFB.}
    \label{fig:G_1}
    \end{subfigure}
    
   \begin{subfigure}[b]{0.9\textwidth} \includegraphics[width=0.9\textwidth]{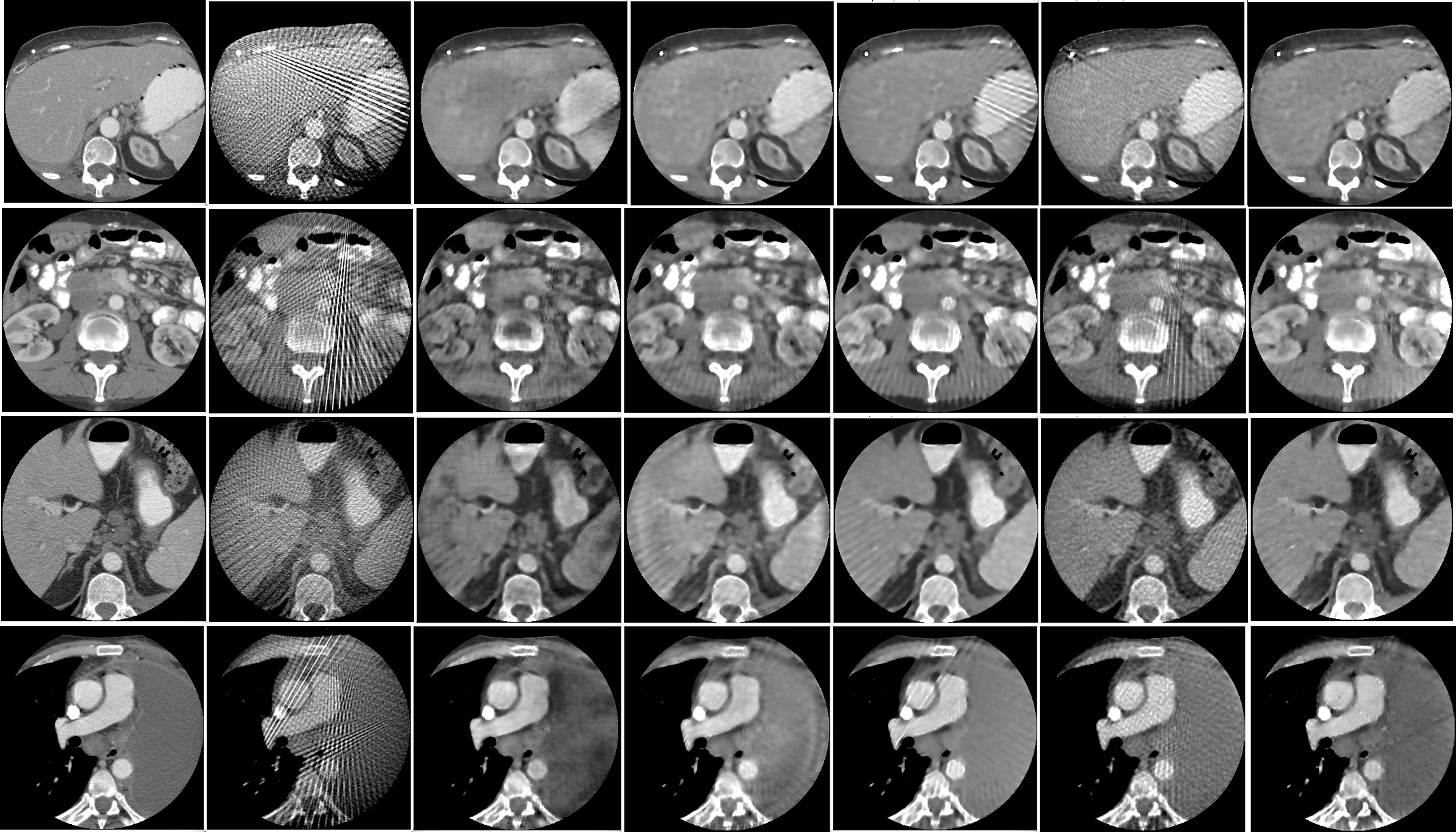}
    \caption{Reconstructed ROIs using different deep learning methods on four examples in the test set of the Abdomen dataset. From left to right: $\overline{\xb}_{\rm ROI}$, FBP, U-net, PNN, ISTA-net, PD-net, U-RDBFB}
    \label{fig:ROI_1}
    \end{subfigure}
    
    \caption{Evaluation on the Abdomen dataset.}
\end{sidewaysfigure*}

Table \ref{tab:1} reports the performance of U-RDBFB compared to U-net and other deep unfolding networks on the testing set of the Abdomen dataset and Figures~\ref{fig:loss1}-\ref{fig:loss2} display the training and testing losses as a function of the number of epochs for all these networks. U-RDBFB performs, on average, better than the other unfolding networks (PNN, PD-net, and ISTA-net) and U-net for all considered metrics. We note that the peaks in the training and testing losses associated with U-RDBFB correspond to the addition of a new data layer during incremental training.
\\
Figure \ref{fig:ROI_1} illustrates the reconstructed ROIs for four examples from the test set of the Abdomen dataset. The FBP reconstruction is also displayed as it is also the input of U-net. The figure confirms that U-RDBFB reduces streaks more effectively than the other unfolding networks. 
At first sight, the images produced by U-net have fewer artifacts than most deep unfolding networks. However, in the second-row and fourth-row images, U-net introduces an artificial dark background. This observation highlights that U-net can hallucinate structures under the sub-sampling streaks of the {FBP} input. Unfolding networks avoid these hallucinations; by simply alternating between U-net and several consistency layers, PNN already minimizes this effect. \\
Figure~\ref{fig:G_1} shows the complete reconstruction on grid $G$ for all unfolding networks. In all cases, since the training loss acts on the ROI only, the exterior is always poorly reconstructed (with ISTA-net, it is very sparse). 
{
\begin{table*}[t]
\centering
\begin{tabular}{|c|c|c|c|c|c|c|}
\hline
\textbf{Metrics}   &  U-RDBFB
& U-net & PPN  & PD-net & ISTA-net & RDBFB\\ \hline
\multicolumn{1}{|l|}{\textbf{PSNR}}   &\textbf{38.7}  &{38.6}     &{\textbf{38.7}}      & 36.5  &{38.2}    &33.9    \\ \hline
\multicolumn{1}{|l|}{\textbf{SSIM}}     &\textbf{0.981}    &{0.972 }   &{0.975}     &{0.956}      &{0.975}     &0.903  
\\ \hline
\multicolumn{1}{|l|}{\textbf{MAE} ($\times 10^{-3}$)}    &\textbf{3.54}  &{4.80}   &{4.00}     &5.53     &{4.95}  &8.41    \\ \hline
\multicolumn{1}{|l|}{\textbf{PieApp}}  &\textbf{0.389}    &{0.502}     &{0.611}     &0.599  &{0.613}   &0.653   \\ \hline
\end{tabular}
\caption{Quantitative assessment of the reconstructed ROIs. Mean values computed over the test set of the Abdomen dataset.}
\label{tab:1}
\end{table*}}
\begin{table*}[t]
\centering
\begin{tabular}{|c|c|c|c|c|c|c|}
\hline
\textbf{Metrics}    &  U-RDBFB
& U-net & PPN  & PD-net & ISTA-net & RDBFB\\ \hline
\multicolumn{1}{|l|}{\textbf{PSNR}}  &\textbf{31.7}  &{16.8}      &{18.5}       &{17.6}   &{23.1}     &21.1    \\ \hline
\multicolumn{1}{|l|}{\textbf{SSIM}}     &\textbf{0.979}    &{0.902}     &{0.838}      &0.765      &{0.957}     & 0.908  \\ \hline
\multicolumn{1}{|l|}{\textbf{MAE} ($\times 10^{-3}$)}     &\textbf{2.17}  &{7.86}    &{4.87}      &6.47     &{3.71}   &6.25    \\ \hline
\multicolumn{1}{|l|}{\textbf{PieApp}}   &\textbf{0.276}    &{0.885}    &{0.962}     &0.984    &{0.650}    &0.476   \\ \hline
\end{tabular}
\caption{Quantitative assessment of the reconstructed ROIs. Mean values computed over the test set of the Head dataset.}
\label{tab:2}
\end{table*}
\begin{table*}[t]
\centering
\begin{tabular}{|c|c|c|c|c|c|c|}
\hline
\textbf{Metrics}    &  U-RDBFB
& U-net & PPN  & PD-net & ISTA-net & RDBFB\\ \hline
\multicolumn{1}{|l|}{\textbf{PSNR}}  &\textbf{27.3}  &{25.1}      &{27.0}       &25.2   &{25.8}     &26.1    \\ \hline
\multicolumn{1}{|l|}{\textbf{SSIM}}     &\textbf{0.856}    &{0.653}      &{0.733}      &0.736   &{0.816}  & 0.848  \\ \hline
\multicolumn{1}{|l|}{\textbf{MAE} ($\times 10^{-3}$)}     &16.6  &{44.1}    &{52.5}      &25.4     &{18.4}   &\textbf{16.5}    \\ \hline
\multicolumn{1}{|l|}{\textbf{PieApp}}   &\textbf{0.267}    &{1.170}     &{1.265}     &1.324    &{0.897}    &0.158   \\ \hline
\end{tabular}
\caption{Quantitative assessment of the reconstructed ROIs. Mean values computed over the testing set of the Geometrical dataset.}
\label{tab:3}
\end{table*}

\subsection{Changing the testing set}

\begin{sidewaysfigure*}[htbp]
\centering
    
     \begin{subfigure}[b]{0.9\textwidth}
    \centering
         \includegraphics[width=\textwidth]{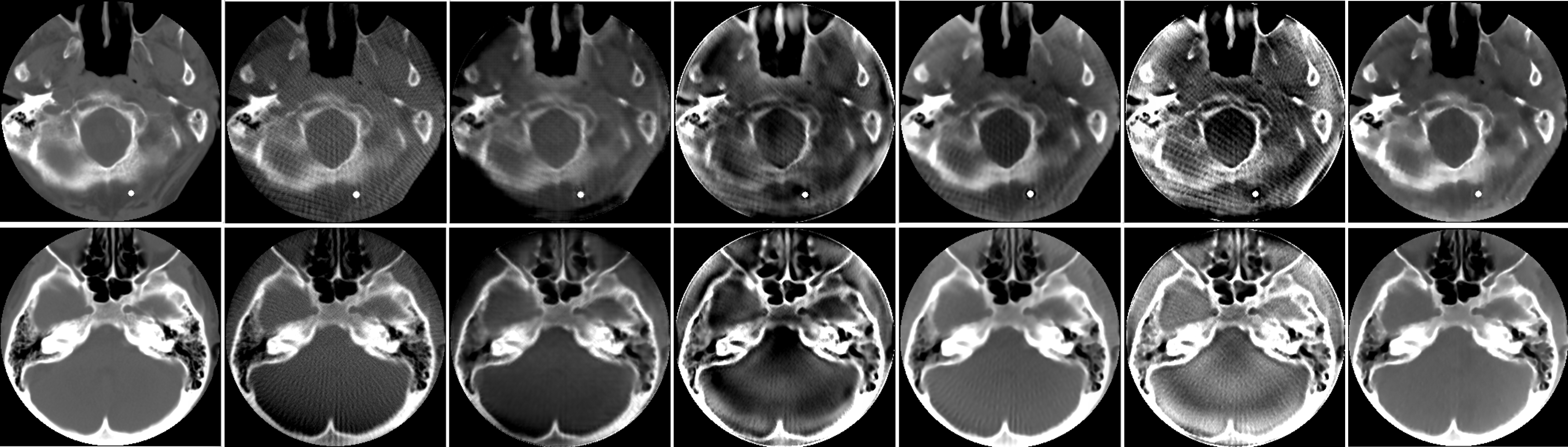}
    \caption{Reconstructed ROIs using different deep learning methods on four examples in the test set of the Head dataset. From left to right: $\overline{\xb}_{\rm ROI}$, FBP, U-net, PNN, ISTA-net, PD-net, U-RDBFB}
    \label{fig:ROI_2}
    \end{subfigure}
\\
   \begin{subfigure}[b]{0.9\textwidth}
   \centering\includegraphics[width=\textwidth]{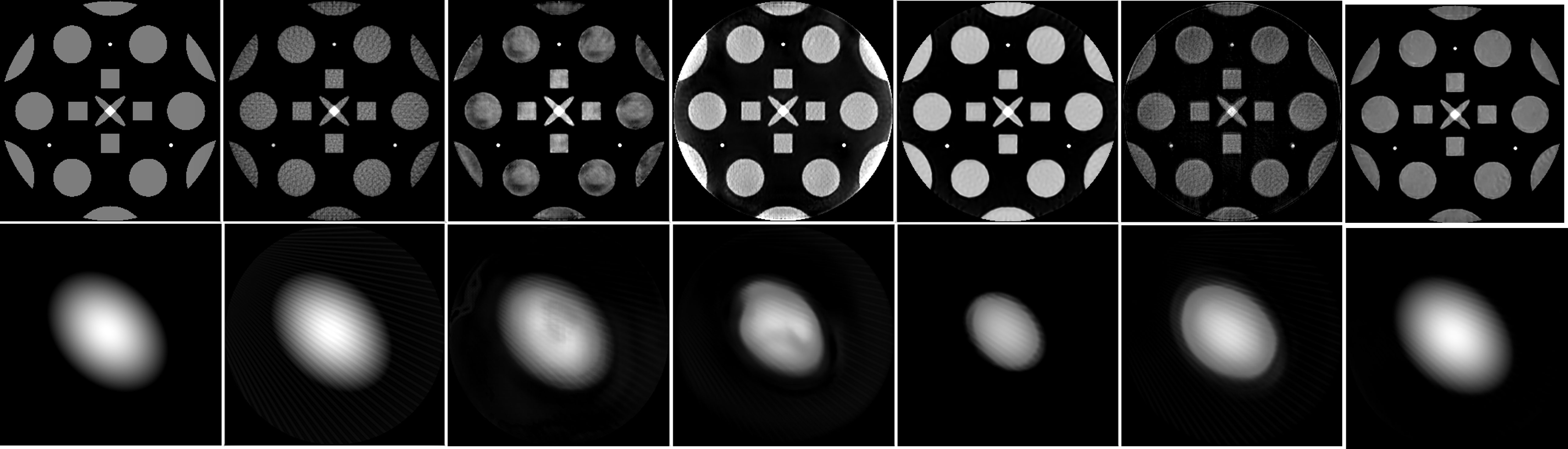}
    \caption{Reconstructed ROIs using different deep learning methods on four examples in the test set of the Geometrical dataset. From left to right: $\overline{\xb}_{\rm ROI}$, FBP, U-net, PNN, ISTA-net, PD-net, U-RDBFB}
    \label{fig:ROI_3}
    \end{subfigure}
    \caption{Evaluation on the Head and Geometrical datasets.}
\end{sidewaysfigure*}

We now evaluate the generalization ability of the methods trained on the Abdomen dataset, using examples from the Head and Geometrical datasets. In Table~\ref{tab:2} and Table~\ref{tab:3}, we report the performance of the trained networks when tested on the Head and the Geometrical datasets. U-RDBFB outperforms the competing unfolding networks for both datasets. ISTA-net is second-best on the Head dataset, and the iterative algorithm ranks second-best on the Geometrical dataset. \\
The four reconstructed images displayed in Figures \ref{fig:ROI_2}-\ref{fig:ROI_3} confirm this trend.
For the Head dataset, U-net performs poorly relative to U-RDBFB on all metrics except for SSIM, where the two methods are rather close. One explanation is that it introduces an offset in some images while limiting the streaks (cf second head image). However, when applied to the geometrical images, it yields unwanted background patterns that strongly degrade our metrics. Offsets and background artifacts are also visible with most unfolding networks, especially PNN and PD-net, except for U-RDBFB. We still note that the head images and the first geometrical image obtained with U-RDBFB have a slightly patchy look, often characteristic of TV regularization. This suggests that U-RDBFB retains the characteristics of the original optimization problem, avoiding the generation of unexpected content as is possible with U-net.

\section{Discussion}\label{sec:disc}

The results show that U-RDBFB outperforms its iterative counterpart as measured in PSNR, SSIM, and PieApp. It not only presents results similar to its iterative counterpart RDBFB in terms of streaks reduction, but it also recovers image details in a much lower number of iterations. We note that U-RDBFB leads to smoother images: the noise is reduced, but the resolution of the image is somehow decreased. This could be explained by the fact that U-RDBFB is designed to minimize the $\ell_2$ norm of the error with respect to the noiseless ground truth, using very few regularization layers; it tends to selectively smooth some parts of the image to remove remaining artifacts. 
\\
All metrics agree that U-RDBFB improves upon learned post-processing U-net and other unfolding networks for our Abdomen dataset. U-net was often associated with a high PSNR compared to the other reconstructions, but this was not always reflected by the PieApp metric. This may be explained by the learned post-processing being limited by the information content of the FBP input, while the unfolding networks act directly with the information content of the data, which is greater than that of the FBP. \\
The computation time for U-RDBFB was about 200 ms in GPU for a $400 \times 400$ reconstruction grid. This is much faster than the iterative reconstruction, which, in our case, requires around 180 s after the regularization parameters have been selected, but slower than other deep learning methods (38 ms for U-net, 68 ms for PNN, 74 ms for ISTA-net, 85 ms for PD-net).\\
U-RDBFB contains fewer learnable parameters than all the other networks. Thanks to our incremental strategy, training U-RDBFB was also found to be easier than other unfolding networks, such as PNN, whose stability highly depends on the initialization for parameter $\lambda$ and the learning rate. Optimizing the architecture and, more precisely, the number of parameters of a neural network is key to transferring its performance to out-of-distribution examples, as shown on the Head and Geometrical datasets. 
Generally, deep unfolding networks are introduced mainly to ensure data consistency through $\Hb$ and embed a fast optimization scheme for fast inference. Our results suggest that including additional a priori knowledge can further boost the performance of deep-learning-based techniques. Note that the structure of U-RDBFB ($K$, $N_k$ and distribution of $\mathcal{L}_D$ and $\mathcal{L}_R$) was not extensively fine-tuned. Our results illustrate that the most straightforward choices work well in our context of ROI imaging from angularly sub-sampled data. \\
Our results also hinted that even without learning, trading a quadratic fidelity for a Cauchy fidelity, discarding the data incompatible with the a priori support of the object, and including some preconditioning through the ramp filter are still of interest to improve reconstruction and reduce the number of iterations.

\section{Conclusion}\label{sec:ccl}
In this paper, we introduced an iterative reweighted algorithm where each inner optimization problem is solved using dual block coordinate forward-backward iterations, and we proposed an unfolded version of it, yielding a neural network for ROI reconstruction from a few measurements.
These methods include a convex surrogate to a Cauchy data fidelity and a TV-based regularization to limit sub-sampling streaks originating from inside and outside the reconstruction grid.
Our experiments demonstrated the benefits of the unfolded strategy over the original iterative algorithm. By balancing the capacity of the network and the use of prior knowledge, our architecture displayed high generalization ability compared to various neural networks, including U-net and other deep unfolding networks.
Future work will investigate loss functions that better preserve image resolution and further optimization of the structure of our network for application to other geometries (e.g., cone-beam) using real CT data.

\appendices
\section{Reweighting strategy}\label{a:MM}
Let $\phi$ be given by \eqref{eq:Cauchy_func}. It was shown in \cite{chouzenoux2013majorize} that, for every $\overline{\zeta}\in\RR$, the following convex quadratic function $\tilde{\phi}(\cdot, \overline{\zeta})$, defined for every 
$\zeta \in \RR$ as
\begin{equation}
    \tilde{\phi}(\zeta , \overline{\zeta}) = \phi(\overline{\zeta}) +   \beta \frac{(\zeta-\overline{\zeta})\overline{\zeta}}{1+(\overline{\zeta}/\kappa)^2} +  \frac{\beta}{2} \frac{(\zeta-\overline{\zeta})^2}{(1+(\overline{\zeta}/\kappa)^2)},
\end{equation}
is a tangent majorant approximation to $\phi$ at $\overline{\zeta}$, that is
\begin{equation}
(\forall \zeta \in \RR)\;\;
    \tilde{\phi}(\zeta , \overline{\zeta}) \geq \phi(\zeta) \quad \text{ and }  \quad \tilde{\phi}(\overline{\zeta} , \overline{\zeta}) = \phi(\overline{\zeta}).
\end{equation}

This allows us to deduce a tangent majorant function $\tilde{g}$ of function $g$ at any point $\overline{\zb} \in \RR^T$: $(\forall \zb \in \RR^T)$,
\begin{align*}
    &\tilde{g}(\zb, \overline{\zb} ) 
    = \sum_{t=1}^T \phi(\zb_t, \overline{\zb}_t) \nonumber\\
    &= g(\overline{\zb}) +  \beta  \operatorname{diag}\left(\left(\frac{\overline{\zb}_t }{1 + (\overline{\zb}_t /\kappa)^2}\right)_{t=1}^T\right) (\zb -\overline{\zb}) \nonumber\\
    &+   \frac{\beta}{2} (\zb -\overline{\zb})^\top \operatorname{diag}\left(\left(\frac{1}{1 + (\overline{\zb}_t /\kappa)^2}\right)_{t=1}^T\right) (\zb -\overline{\zb}) \geq g(\zb ).
\end{align*}
Finally, for every $\overline{\xb} \in \RR^L$, we set
\begin{equation}
(\forall \xb \in \RR^L)\quad
    \tilde{f}(\xb,\overline{\xb}) = \tilde{g}(\Hb \xb - \yb;\Hb \overline{\xb} - \yb), 
\end{equation}
that satisfies $\tilde{f}(\xb, \overline{\xb}) \geq g(\Hb \xb - \yb) = f(\xb)$.
Given this majoration, the iterative reweighting strategy approximates the solution to \eqref{eq:pb0} by the estimate produced by Algorithm \ref{alg:rewighted}, where
\begin{align}\label{eq:pb0_R}
    (\forall (\xb,\overline{\xb}) \in (\RR^L)^2) \quad Q(\xb, \overline{\xb}) = &\tilde{f}(\xb , \overline{\xb} ) + r(\xb). 
\end{align}

\section{Implementation of DBFB}\label{a:DFB}
Step \textbf{(D)} involves the calculation of the proximity operator $\prox_{\gamma_n^{-1} \sigma h_0 (\cdot, \Bb_0 \overline{\xb})}$, which has a closed-form \cite[Example 24.2]{Livre1}, for
$(\forall (\zb, \overline{\zb}) \in (\RR^T)^2)$,
\begin{align}
    &\prox_{\gamma_n^{-1} \sigma  h_0(\cdot , \Bb_0 \overline{\zb})}(\zb) = \prox_{\gamma_n^{-1} \sigma \tilde{g}(\cdot -  \yb, \overline{\zb} -  \yb)}(\zb), \nonumber \\
        &=  \yb + \prox_{\gamma_n^{-1} \sigma \tilde{g}(\cdot, \overline{\zb}- \yb)}(\zb -  \yb)\nonumber\\
    &= \left( \yb_t + \frac{\zb_t - \yb_t}{{1+}\beta \gamma_n^{-1} \sigma \big(1+ (\overline{\zb}_t -  \yb_t)^2/\kappa^2\big)^{-1}} \right)_{t=1}^T. 
    \label{eq:proxh0}
\end{align}
Step \textbf{(R)} requires calculating the proximity operator of $h_1$ scaled by parameter $\gamma \in \RPP$. It also has a closed form: for $\sb = (\sb_1,\ldots,\sb_J) \in \RR^{2JL}$, $\prox_{\gamma h_1}(\sb) = \left(\prox_{\gamma r_j}(\sb_j)\right)_{j=1}^{J}$,
where, for every $\zb = (\zb_1,\zb_2) \in \RR^{2L}$,
\begin{equation}\label{e:proxgahj}
\prox_{\gamma r_j }(\zb) = \left(
\max\Big\{0,1-\frac{\gamma  \alpha_{j,l}}{
\|\zb_{\ell}\|_2}\Big\} \zb_{\ell}
\right)_{\ell=1}^L,
\end{equation}
where, for every $\ell \in \{1,\ldots,L\}$, $\zb_\ell = ((\zb_1)_{\ell},(\zb_2)_{\ell}) \in \RR^2$.

\section{Parameter learning}\label{a:PL}

We discuss our choices for $(\mathcal{L}_{k,n}^{(\theta)})_{n=0}^{N_k-1}$.

\begin{itemize}
    \item step size for $\mathcal{L}_{\rm D}$:  $\nu_{k,n,0} = \mathcal{L}_{k,n}^{(\nu)} =  \operatorname{softplus}(a_{k,n})$ where $a_{k,n}$ is a learnable real-valued parameter.
    \item Parameters of the Cauchy function for $\mathcal{L}_{\rm D}$:
    \begin{itemize}
    \item[$*$] $\kappa_{k,n} = \mathcal{L}_{k,n}^{(\kappa)}  = W_{\kappa} \operatorname{softplus}(c_{k,n})$ where $c_{k,n}$ is inferred from a fully connected layer, whose weights are shared across the U-RDBFB network, applied on a histogram of the absolute value of the filtered reprojection error, i.e., $\Hb^\top (\Fb \xb_n-\yb)$. We implemented the learnable histogram layer proposed in \cite{Wang2016LearnableHS}, which is piecewise differentiable. More precisely, we built a cumulated histogram using 100 bins from 0 to the maximum value of the filtered reprojection error. 
     \item[$*$] $\beta_{k,n} = \mathcal{L}_{k,n}^{(\beta)} = W_{\beta} \operatorname{softplus}(d_{k,n})$.
    \end{itemize}
     \item Diagonal elements of $\Mb$ involved in \eqref{eq:expression_r} corresponding to the locations of pixels outside of the ROI for both $\mathcal{L}_{\rm D}$ and $\mathcal{L}_{\rm R}$:
     $\xi_{k,n} = \mathcal{L}_{k,n}^{(\xi)} =  \operatorname{softplus}(e_{k,n})$ where $e_{k,n}$ is learned.
        \item step size for $\mathcal{L}_{\rm R}$:  
            For every $j \in \{1, \ldots, J\}$, $\nu_{k,n,j} = \mathcal{L}_{k,n}^{(\nu_j)} =  W_{\nu} \operatorname{softplus}(b_{k,n,j})$. 
        \item Parameters of the STV regularization for $\mathcal{L}_{\rm R}$: For every $j \in \{1, \ldots, J\}$,
    \begin{align*}        \alpha_{k,n,j} & = (\alpha_{k,n,j,l})_{\ell=1}^L
    = \mathcal{L}_{k,n}^{(\alpha_{j})}
    \nonumber\\
    &= 
    W_{\alpha} \operatorname{softplus}( A_{k,n} \circ \operatorname{relu} \circ B_{k,n}(\nabla_j \xb_k)),  
    \end{align*}
    where $A_{k,n}$ is a grouped convolution of 7 groups with size $3 \times 3$ kernels and $J=7$ channels and $B_{k,n}$ is a grouped convolution of 14 groups with size $5 \times 5$ kernels and $14$ channels.
     \end{itemize}

In $\mathcal{L}_{D}$, initial values of $a_{k,n}$, $d_{k,n}$ are set to $1$.
In $\mathcal{L}_{R}$, initial values for $b_{k,n,j}$, $e_{k,n}$ are $1$.
Normalization scalars $W_{\kappa}$, $W_{\nu}$, $W_{\beta}$ and $W_{\alpha}$ are set to $10^{-5}$, 10, $10$ and $0.05$ respectively.
\\
Finally, in each layer $\mathcal{L}_{R}$, operations $(\tilde{\mathbf{L}}^{k,n}_j)_{j=1}^J$ are learned; they are convolutions with a kernel of the same size as for $(\mathbf{L}^{k,n}_j)_{j=1}^J$.

\end{document}